\documentclass[useAMS,usenatbib,usegraphicx]{mn2e}

\def\vism{$v_{\rm ISM}$}

\def\kms{$\mathrm {km\,s}^{-1}$}
\def\mdot{$\mathrm {M_{\odot}\,yr}^{-1}$}

\def\density{$10^{-6}\ \mathrm {M_{\odot}\,pc}^{-3}$}

\def\hcm3{$\mathrm{H\,cm}^{-3}$}

\def\oversim#1#2{\lower0.5pt\vbox{\baselineskip0pt \lineskip-0.5pt
     \ialign{$\mathsurround0pt #1\hfil##\hfil$\crcr#2\crcr\sim\crcr}}}

\title[AGB/PN -- ISM interaction]{The interaction of planetary nebulae 
and their AGB progenitors with the interstellar medium}
\author[C. J. Wareing et al.]
{C. J. Wareing$^{1,2}$\thanks{E-mail: cjw@maths.leeds.ac.uk}, 
Albert A. Zijlstra$^{1}$\thanks{E-mail: a.zijlstra@manchester.ac.uk},
T. J. O'Brien$^{1}$\thanks{E-mail: tim.obrien@manchester.ac.uk}\\
$^{1}$Jodrell Bank Centre for Astrophysics, Alan Turing Building, 
The University of Manchester, Oxford Road, \\Manchester, M13 9PL, UK\\
$^{2}$Department of Applied Mathematics, University of Leeds, Leeds,
LS2 9JT, UK}

\begin{document}

\date{}


\maketitle

\label{firstpage}

\begin{abstract}

Interaction with the Interstellar Medium (ISM) cannot be ignored in
understanding planetary nebula (PN) evolution and shaping.  In an effort to
understand the range of shapes observed in the outer envelopes of PNe, we have
run a comprehensive set of three-dimensional hydrodynamic simulations, from
the beginning of the asymptotic giant branch (AGB) superwind phase until the
end of the post--AGB/PN phase. A 'triple-wind' model is used, including a slow
AGB wind, fast post--AGB wind and third wind reflecting the linear movement
through the ISM.  A wide range of stellar velocities, mass-loss rates and ISM
densities have been considered.

We find ISM interaction strongly affects outer PN structures, with 
the dominant shaping occuring during the AGB phase. The simulations 
predict four stages of PN--ISM interaction whereby the PN is initially
unaffected (1), then limb-brightened in the direction of motion (2), then
distorted with the star moving away from the geometric centre (3) and
finally so distorted that the object is no longer recognisable as a 
PN and may not be classed as such (4). Parsec-size shells around 
PN are predicted to be common. The structure and 
brightness of ancient PNe is largely determined by the ISM interaction, 
caused by rebrightening during the second stage; this effect may
address the current discrepancies in Galactic PN abundance. The 
majority of PNe will have tail structures. Evidence for strong interaction is
found for all known planetary nebulae in globular clusters.

\end{abstract}


\section{Introduction}

Planetary nebulae (PNe) display a wide variety of shapes ranging from round,
which can be simply understood in terms of the symmetric interacting stellar
winds (ISW) model \citep{kwok82}, through complex symmetrical shapes, such as
hour-glasses and butterflies, to shapes which have only rotational
point-symmetry. Many theories have been introduced to explain these complex
shapes from adding an asymmetric slow wind to the ISW model
\citep{kahn85,balick87} to involving the effects of binary central stars
\citep{soker96} and magnetic fields \citep{frank04}. Observations of PNe have
shown several cases where only the outer shell shows a departure from
symmetry. In these cases, the cause of the asymmetries has been postulated to
be an interaction with the interstellar medium (ISM).

Interaction with the ISM by PNe was first discussed by \cite{gurzadyan69} with
an early theoretical study by \cite{smith76} employing the 'snow-plough' model
of \cite{oort51}. Smith concluded that a nebula will fade away before any
disruption of the nebular shell becomes noticable. \cite{isaacmann79} used the
same approximation with higher relative velocities to the ISM and ISM densities and
concluded similarly.

In contrast, \cite{borkowski90} found that many PNe with large angular extent
show signs of PN--ISM interaction and that all nebulae containing central
stars with a proper motion greater than 0.015 arcsec\,yr$^{-1}$ do so.
\cite{soker91} hydrodynamically modelled the interaction revealing that the PN
shell is first compressed in the direction of motion and then in later stages
this part of the shell is significantly decelerated with respect to the
central star.  Both conclude that the interaction with the ISM becomes
dominant when the density of the nebular shell has dropped below a certain
critical limit, typically $n_{\rm H}=40$\,cm$^{-3}$ for a PN in the Galactic
plane.  These low densities require large, evolved nebulae, in agreement with
the observational result of \cite{borkowski90}. They noted that their simple
picture breaks down for high velocity PNe in a low density environment. Here,
a Rayleigh--Taylor (RT) instability develops, leading to shell fragmentation. Their
2D hydrodynamic simulations started with the nebula shell already formed but
above their upper density limit for ISM interaction to become apparent.

The fragmentation predicted by \cite{soker91} was discussed in more depth by
\cite{dgani94} and then \cite{dgani98}. In these papers, Dgani and Soker
applied theoretical results of hydrodynamic instabilities and found that the
RT instability can play an important role in the shaping of the
outer regions of a PN. They suggested these RT instabilities can cause
fragmentation of the bow shock with Kelvin--Helmholtz instabilities also
playing a part. Any fragmentation caused by these instabilities would only be
present if the relative velocity to the ISM of the central star was greater than 100
\kms.

\cite{villaver03} (hereafter referred to as VGM) pointed out that the PN--ISM
interaction had previously been studied by considering the interaction after
the nebular shell had formed.  PNe are formed when a slow wind ($\sim10$ \kms)
ejected during the preceding asymptotic giant branch (AGB) phase of evolution
is swept up into a dense shell by a fast wind ($\sim10^3$ \kms) from the
exposed, ionizing hot white dwarf core.  VGM performed 2D hydrodynamic
simulations and found that crucially the interaction begins during the AGB
phase where the slow wind is shaped by the ISM. The PN forms in this
pre-shaped environment. Choosing a conservative relative velocity of the
central star to the ISM of 20 \kms\ and a low density of the surrounding ISM of $n_{\rm
H}=0.1$ cm$^{-3}$, they discovered that the PN is brightened on the upstream
side of the nebular shell. They concluded that PN--ISM interaction provides an
adequate mechanism to explain the high rate of observed asymmetries in the
external shells of PNe. Further, stripping of mass downstream during the AGB
phase provides a possible solution to the problem of missing mass in PN
whereby only a small fraction of the mass ejected during the AGB phase is
inferred to be present during the post--AGB phase. Observational evidence for
the effect of the ISM on AGB wind stuctures, supporting VGM's findings was
found by \cite{zijlstra02}.

VGM found that simple hydrodynamic simulations can reveal much 
information regarding the PN--ISM interaction. In order to investigate the
interaction further, we have developed a 'triple-wind' model including
an initial slow AGB wind, a subsequent fast post--AGB wind, and a third 
continuous wind reflecting the movement through the ISM. Employing a parallel 3D
hydrodynamic scheme developed by \cite{wareing05}, we have understood 
the formation of the extreme PN Sh 2-188 \citep{wareing06a} and the 
structure around the AGB star R Hya \citep{wareing06b}. 

Sh 2-188 was thought to be a bright one-sided arc-like PN when 
new observations taken as part of the Isaac Newton Group Photometric 
H$\alpha$ Survey of the Northern Galactic Plane (IPHAS) \citep{drew05} 
revealed a faint ring-like completion of the arc and a tail stretching 
away in opposition to the bright arc. Our model revealed
this PN to be a strong PN--ISM interaction where the central star is moving
at 125 \kms\ in the direction of the bright arc relative to the ISM 
and the nebular shell is 
interacting with a bow shock formed during the AGB phase between the slow
wind and the ISM. Recent IR observations of the AGB star R Hya as part 
of the MIRIAD programme \citep{ueta06} have revealed the arc-like structure 
to the North West of the star to be a bow shock ahead of the star 
\citep{wareing06b}. The existence of this AGB wind bow shock has confirmed the 
hypothesis that the major shaping effect for the PN--ISM interaction 
occurs during the AGB phase of evolution. 

Using our model, we have now run a comprehensive set of 92 simulations
equivalent to over 5 years of single CPU computation time investigating the
PN--ISM interaction. In this paper, we discuss a representative set of these
simulations and generalise the interaction into four distinct stages,
indicating its effect on shaping and other PN characteristics. We apply our
generalisation to a selection of PNe from the IAC Morphological Catalog of
Northern Galactic Planetary Nebulae \citep{manchado96} and find indications of
interaction with the ISM in approximately 20 per cent of objects.

\section{The hydrodynamic scheme and triple-wind model}

The numerical scheme, CUBEMPI, used in our simulations to solve the
hydrodynamics equations employs a second-order Godunov solution due to
\cite{falle91}.  In recent years, variants of CUBEMPI have been used to shed
light on nova explosions \citep{lloyd97,porter98}, extragalactic jet-cloud
interactions \citep{higgins99} and most recently PN \citep{mitchell07,
wareing06a}.  The version of the scheme used here 
is posed in 3D cartesian coordinates, fully parallel and
includes the effect of radiative cooling due to the cooling curves of 
\cite{raymond76} above 10$^4$ K as the cooling curves extend no 
further. The parallelisation was developed using the
MPI\footnote{http://www-unix.mcs.anl.gov/mpi/} library and involves slicing
the numerical domain along an axis and communicating relevant boundary data at
the correct points during a computational timestep. The parallelisation has
been successfully tested for efficiency and scalability.
The scheme itself has also been tested on a number of standard computational 
fluid dynamics problems and performed well on all tests. 
It has been further tested using
astrophysical problems which have highlighted its capability for shock
capturing, an important requirement for the modelling of PNe.
The parallelisation and testing are detailed in \cite{wareing05}

The numerical domain consists of a cubic grid, 200 cells along each axis, 
containing a total of $8\times10^6$ uniformly-spaced cells. The grid is sliced
ten times for parallelisation purposes and distributed across ten processors
of the COBRA beowulf-type supercomputer at Jodrell Bank. This level of 
parallel distribution is a comprimise between computational efficiency and
availability of computational resources.

In the 'triple-wind' model, the simulation is performed in the frame of
reference of the star, which is placed at cell coordinates ($50,100,100$).
Mass loss is effected by setting the values of hydrodynamic variables
(density, momentum density and energy density) in a volume-weighted spherical
region of radius 5$\frac{3}{4}$ cells centred on the star. The radius of this
source volume has been chosen by an experimental process balancing production
of the most spherical PN when modelling the stationary ISW model with
spherically symmetric winds (i.e. reducing the pixelation of the cartesian
grid) and avoidance of interference with results of the simulation. The
conditions within the source volume are reset at the beginning of every
computational timestep to drive the ejection of the wind. The wind has been
modelled with a spherically symmetric constant mass-loss rate $\dot{M}$ with
constant velocity $v$ and temperature $T$.  Density in the source volume has
been defined by $\dot{M}/(4 \pi v r^2)$ where $r$ is the physical radial
distance from the central star.  The other hydrodynamic variables are set
accordingly. Simulation of movement through the ISM is achieved by flowing ISM
material in at the ($x=1$) boundary with a velocity vector v$_x$, v$_y$, v$_z$
= ($+v$,0,0). The ISM density and temperature are constant. All other
numerical boundaries have conditions allowing material to flow out of the
domain freely. Gas pressures in the model are calculated assuming an ideal gas
equation of state.

\section{Model parameters}

In our simulations, we have held the following parameter values constant: for
the slow AGB wind a velocity of $v_{\rm sw} = 15$ \kms\ and a temperature of
$T_{\rm sw} = 10^4$ K as the cooling curves extend no further; 
for the fast post--AGB wind a mass-loss rate of
$\dot{M}_{\rm fw} = 5 \times 10^{-8}$ \mdot, a velocity of $v_{\rm fw} = 1000$
\kms\ and a temperature of $T_{\rm fw} = 5 \times 10^4$ K; and for the ISM a
temperature of $T_{\rm ISM} = 8000$ K, characteristic of the warm intercloud
medium \citep{burton88}. These parameter values are typical of observations of
PN conditions.  In the model, the switch between the AGB wind and the post--AGB
wind is instantaneous and occurs after $5\times10^5$ years of AGB evolution,
typical of the duration of this phase. In view of the still considerable
uncertainties on the detailed properties and evolution of these winds, more
detailed temporal variations have not been modelled. We have discussed this further
in later sections.

In our simulations, we have varied three parameters: relative velocity of the
star to the surrounding ISM \vism, slow wind mass-loss rate
$\dot{M}_{\rm sw}$ and ISM density $\rho_{\rm ISM}$.  \cite{binney98}
discussed \vism\ of stars in the Galactic plane: typical thin
disk stars resulting in a PN appear to have a transverse variation on their
galactic rotation velocity of 35 -- 50 \kms; thick disk stars, rarer in the
Galactic plane, have a typical range of 50 -- 75 \kms; halo objects can have a
typical variation greater than 100 \kms.  Older stellar groups, such as PNe,
are characterized by larger velocity dispersions and asymmetric drift
velocities than are younger stellar groups. The average transverse motions of
stars in the solar neighbourhood have been found to be in the range of 20--40
\kms\ with the tail of the distribution up to 130 \kms\ \citep{skuljan99}. For
white dwarfs an average velocity of $67.4 \pm 39.5$ \kms\ has been found by
\cite{sanggak84}.  In terms of Galactic disk PNe, \cite{borkowski90} state 60
\kms\ as an average velocity. For Galactic Bulge PNe, velocities up to 200
\kms\ could be reasonably expected as these stars are dynamically a much
hotter population. Therefore, we have considered a range of \vism\
from 0 \kms, testing the implementation of the triple-wind model
and its ability to simulate a spherical nebula, up to 200 \kms, in 25 \kms\
steps in order to fully cover the range of velocities of PN-forming stars in
the Galaxy.

The typical density of the ISM in the galactic plane is approximately $n_{\rm
H}$ = $\sim2$ cm$^{-3}$ up to a scale height of 100 pc above the plane where
it then begins to drops off exponentially. We have used three constant values
of ISM density $n_{\rm H}$ = 2, 0.1 \& 0.01 cm$^{-3}$ to investigate this
range. These densities are comparable to that of the warm intercloud medium
\citep{burton88}.

Mass-loss rates during the AGB phase of evolution vary between 10$^{-7}$ --
10$^{-6}$ \mdot\ with brief periods of enhanced mass-loss up to 10$^{-5}$
\mdot. As we are using a constant mass-loss rate, we have not modelled 
mass-loss variations. We have used values of 10$^{-7}$, $5\times10^{-7}$,
10$^{-6}$ \& $5\times10^{-6}$ \mdot\ up to \vism\ = 75
\kms. Above this velocity, we have not used a mass-loss rate of 10$^{-6}$
\mdot\ due to time and computational constraints.

\section{Results}

In our initial test cases \vism\ is set to zero. The
forward shock driven by the AGB wind is spherical and expands into a
homogeneous ISM. The shocked ISM material behind the forward shock is of
relatively low density with the contact discontinuity and the reverse shock
expanding spherically symmetrically outward.  The unshocked AGB wind behind
the reverse shock is also spherically symmetric around the central star with
an approximate 1/r$^2$ distribution.  After 500\,000 years, the fast post--AGB
wind is introduced into this distribution driving a shock
which sweeps up the AGB wind material into an expanding shell. The shell has a
relatively high density and a temperature on the scale of 10--20\,000 K.
The stromgren sphere around the central star encompasses the shell 
indicating the material in the shell would be ionized and observable as a
PN. This method of PN formation is the result of interacting stellar winds and
is the premise of the ISW model.

Next, we added the ISM velocity. We display and discuss a representative 
set of five simulations covering various parameter values which serve to 
illustrate the range of PN structures formed in all our simulations. 
Table \ref{parameters} shows the parameter values for these five simulations.
The resulting PN structure is discussed for each case. In our online
supplementary appendix, we have also included snapshots at 
the end of the AGB phase of evolution and parameter values for our 
full set of simulations. We have performed calculations to find
the extent of the stromgren sphere during the PN phase and find that 
in all cases it extends beyond the simulation domain.

In the following figures, the results are illustrated by slices through the 
density data cubes at the position of the central star and 
parallel to the direction of motion. 

\begin{table}
\caption{Input parameters for the five selected simulations discussed
in section 4. Column a) gives the simulation reference; 
column b) \vism; 
column c) the density of the surrounding ISM in $n_{\rm H}\,\rm cm^{-3}$; 
column d) the constant mass-loss rate during 
the AGB phase of evolution; and column e) the physical size of the grid 
along one dimension of the numerical domain.}
\label{parameters}
\begin{center}
\begin{tabular}{lcccc|}
\hline
a) & b) & c) & d) & e) \\
case & \vism & ISM:$n_{\rm H}$ & $\dot{M}_{\rm sw}$ & Grid \\
   & (\kms) & (cm$^{-3}$) & (M$_{\odot}$\,yr$^{-1}$) & (pc)   \\
\hline
A  & 25 & 0.01 & $5\times10^{-7}$ & 23.4 \\
B  & 50 & 2 & $5\times10^{-6}$ & 2.6 \\
C  & 75 & 2 & $5\times10^{-6}$ & 1.75 \\
D  & 100 & 2 & $10^{-7}$ & 1.0 \\
E  & 125 & 2 & $5\times10^{-6}$ & 1.0 \\
\hline
\end{tabular}
\end{center}
\end{table}

\subsection{Case A}

\begin{figure*}
\begin{center}
\includegraphics[angle=0,width=13.5cm]{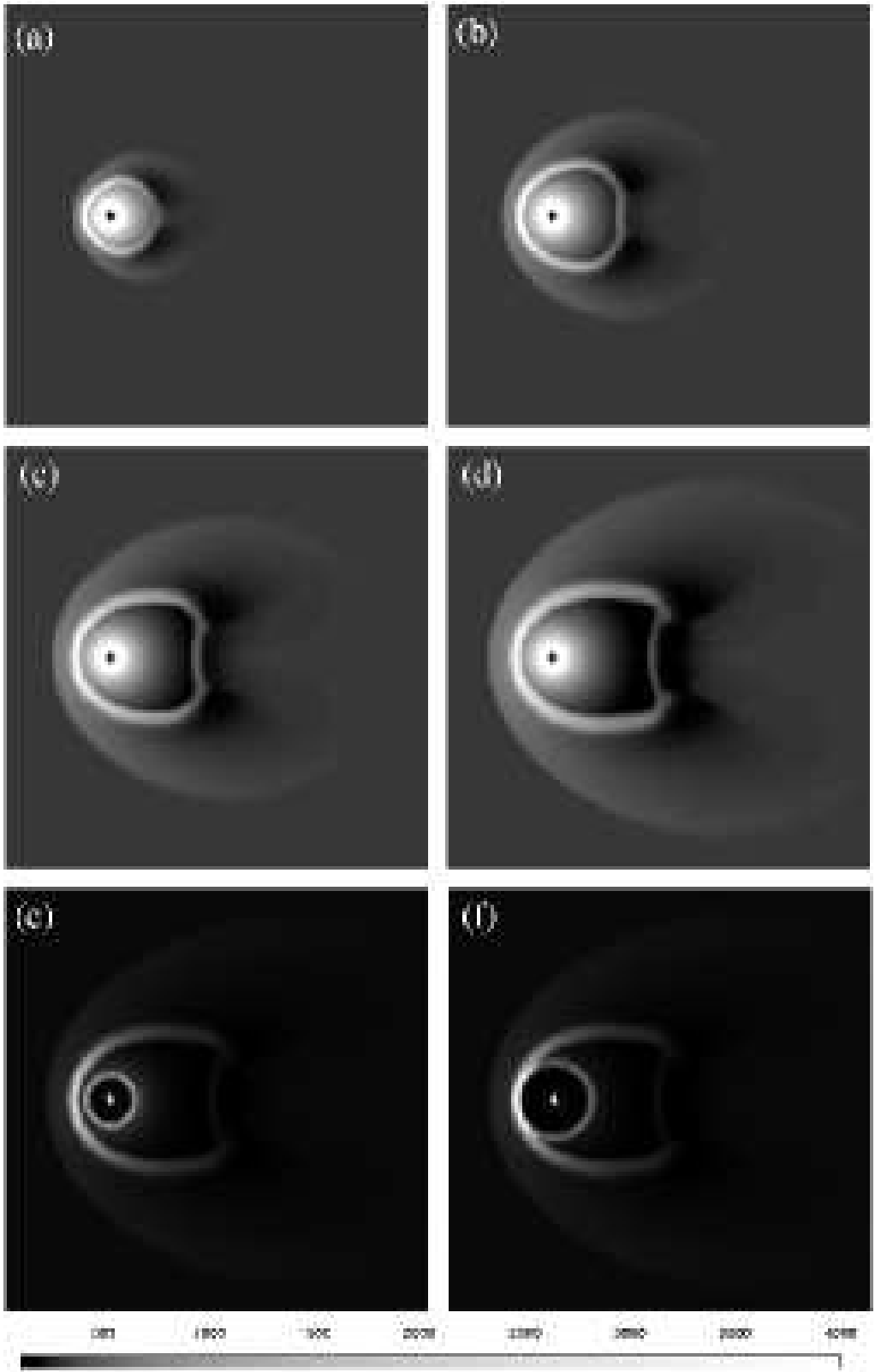}
\caption{The results of case A: the panels show the gas density during
the AGB and post--AGB phases. In panel (a), the simulation is 125\,000
years into the AGB phase, panel (b) 250\,000 years, panel (c) 375\,000
years, panel (d) 500\,000 years, panel (e) 15\,000 years into 
the post--AGB phase and in panel (f) 30\,000 years.
The position of the central star is marked by an asterisk. 
The colour scaling is logarithmic and the density is scaled in units of 
\density\ where 25\,000 is equivalent to $n_{\rm H}$ = 1.0 cm$^{-3}$.}
\label{pn1}
\end{center}
\end{figure*}

Figure \ref{pn1} shows the result of case A. The central star has
\vism\ of 25 \kms. The AGB evolution is shown in the top four
panels (a)--(d) with the post--AGB evolution in the bottom two panels (e) and (f). In
panel (a), 125\,000 years into the AGB phase, the shock has formed into a bow
shock a short distance upstream of the central star with a tail connecting a
short distance downstream of the central star. Over the next three panels (b), (c)
and (d), at 250\,000, 375\,000 and 500\,000 years respectively, the bow shock
can be seen to expand outwards from the central star and between panels (c)
and (d) reach a stable position ahead of the star.  This
can be understood in terms of a ram pressure balance between the slow wind and
the oncoming ISM. At this point, the bow shock is approximately 2.5 pc ahead
of the central star. If we assume this is a strong shock, the temperature of the shocked
material at the head of the bow shock should be equal to (3/16)\,(m\,\vism$^2$/k)
where $m$ is the particle mass in the simulation
and $k$ the Boltzmann constant. Our simulation is in agreement with this. The
tail structure can be understood in terms of material being ram-pressure
stripped from the head of the bow shock and cooling as it flows downstream 
around the bow shock into the tail. It is clear that even at this low 
speed, the ISM interaction strongly affects the shape of the AGB wind.

The brightest phase of the PN evolution, typically 1--5$\times10^3$\,yr into
the post--AGB phase, is not shown. Panel (e) shows the phase 15\,000 years into
the post--AGB phase. The expanding (and now faint) shell of the PN is clear,
inside and still detached from the bow shock of shocked AGB wind material. The PN
and the {\it far older} bow shock are at this time separate objects with the PN
expanding within the bubble of undisturbed AGB wind material; the ISM interaction
has yet to affect the PN. Observations would reveal an ancient symmetric ring
PN approximately 3.5 pc across.  Deeper observations may reveal the cooler
material in the bow shock structure surrounding the PN. In panel (f), 30\,000
years into the PN phase, the PN has now expanded far enough to interact with
the AGB wind bow shock.  The portion of the PN interacting with the bow shock now
rebrightens via the interaction of the bow shock and PN shell. The PN at 
this stage is still circular, the rebrightening being
the first and only indication of an ISM interaction in this case.

\begin{figure}
\begin{center}
\includegraphics[angle=0,width=6.5cm]{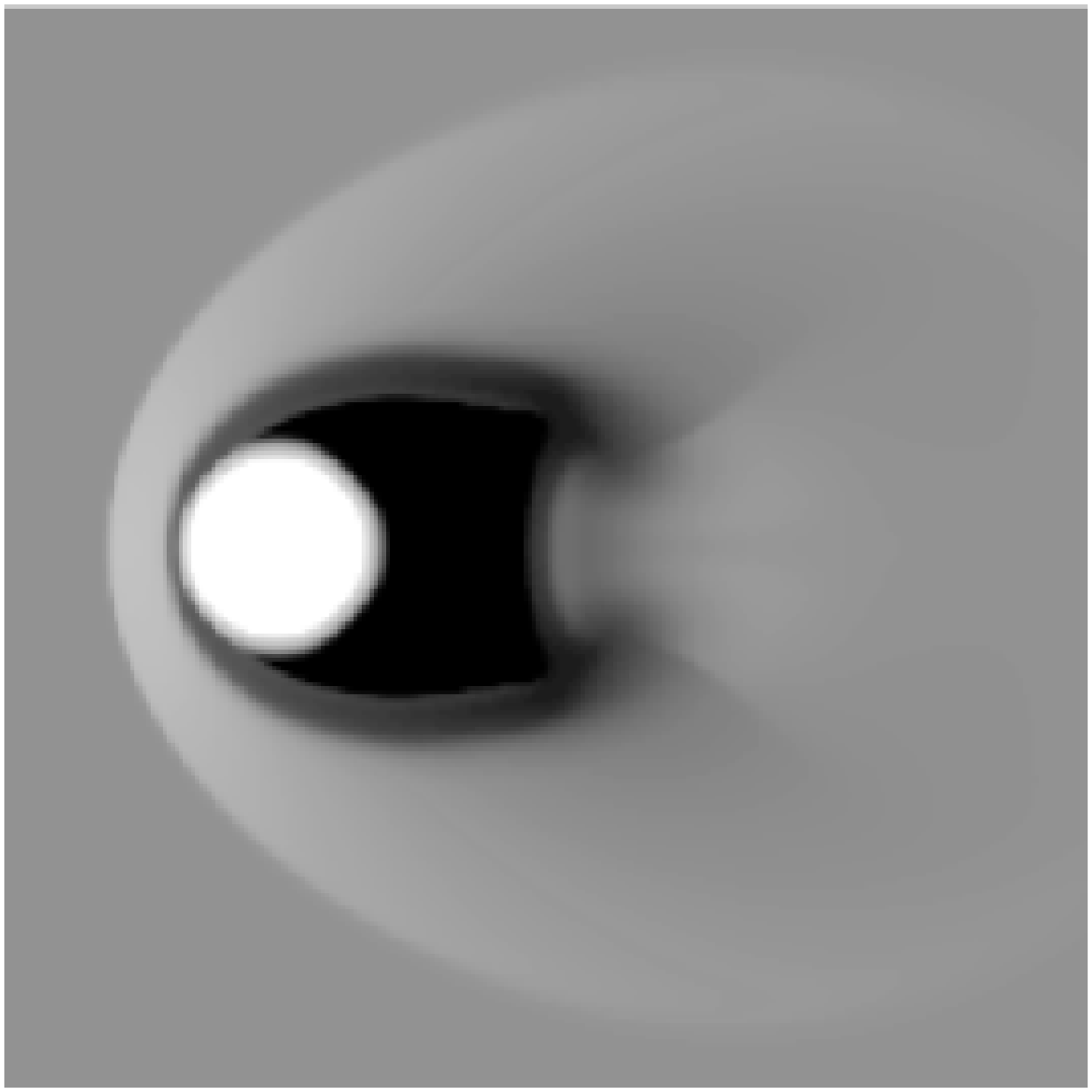}
\includegraphics[angle=0,width=6.5cm]{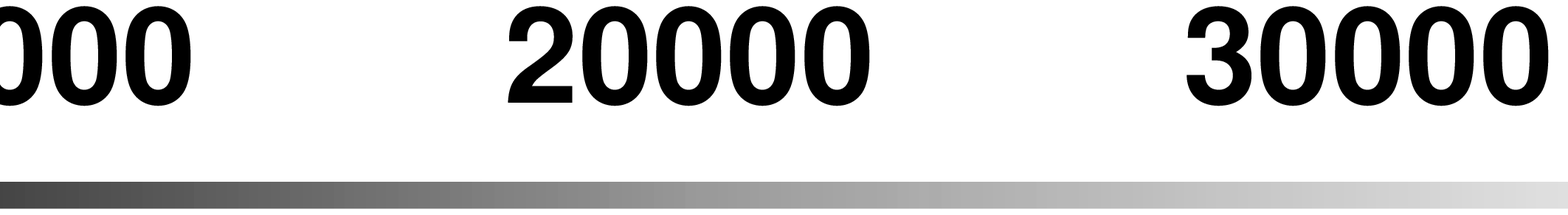}
\caption{In this figure we show the temperature profile (in Kelvin) of 
the 30\,000-year old PN shown in panel (f) of figure \ref{pn1} on the
same slice through the numerical domain.}
\label{pn1-temp}
\end{center}
\end{figure}

Figure \ref{pn1-temp} shows the temperature distribution of the PN shown 
in panel (f) of Figure \ref{pn1}, 30\,000 years into the PN phase.
Given our calculation that the Stromgren sphere extends beyond
the simulation domain, we assume that everything which is at 10$^4$ K or lower
is photo-ionised and that everything which has a temperature greater than
$1.5 \times 10^4$ K is collisionally ionised. Thus the material 
at the head of the bow shock appears to be collisionally ionised. Looking at
the tail, the material is cooler and thus may be photo-ionised if 
the central star is still bright enough. This is the general case for the
rest of our simulations where collisional ionisation of material can be 
inferred at the head of the bow shock with photo-heating/ionisation of the 
material stripped into the tail.

We define the previous phase where the PN was inside the AGB wind bubble as
the first stage of PN--ISM interaction. The second stage would begin with the
first indication of ISM interaction when the PN brightens in the direction of
motion. This second stage will be responsible for rebrightening of 
ancient PNe and including this effect in the projection of current PN Galactic
distributions may address the very long visible life time of PNe implied by
recent studies \citep{moe06, zijlstra91}.

The ISM interaction is clearly important even at these low speeds, as 
it shapes the AGB wind long before the PN phase. The AGB wind forms a bow 
shock upstream of the central star with tails stretching downstream.
The PN does not interact with the AGB wind bow shock until it has expanded 
enough to reach the bow shock, typically after 25--30\,000 years. At that
point, the enhanced density and temperature of the material in the region
of interaction suggests strengthening of the emission from that area:
rebrightening.
The parameter values in this case are comparable to those used by VGM
and we have found similar effects on the PN formed, supporting our
triple-wind model and conclusions.

\subsection{Case B}

\begin{figure*}
\begin{center}
\includegraphics[angle=0,width=14cm]{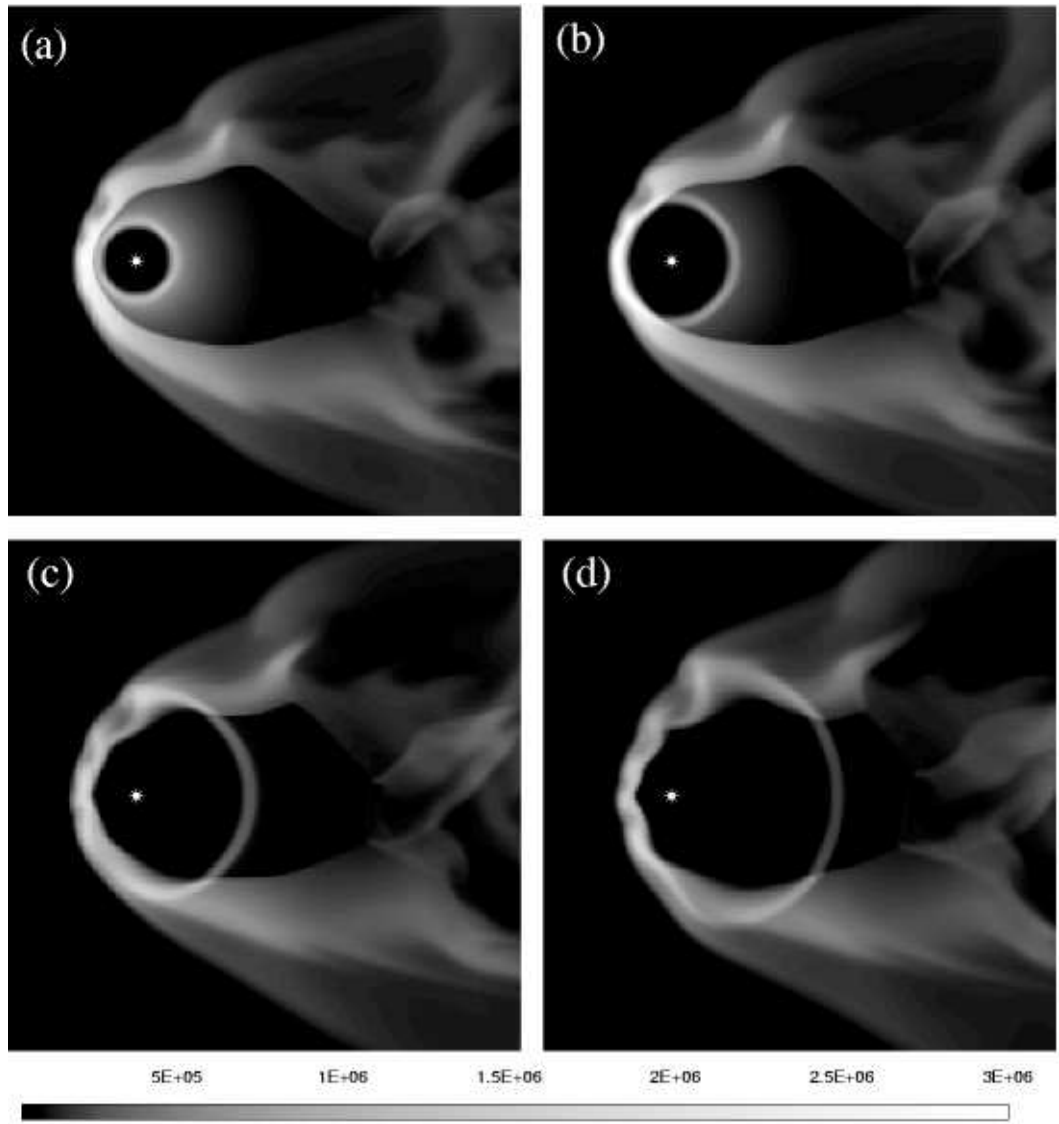}
\caption{The results of case B: the panels show the gas density during
the post--AGB phase. In panel (a), the simulation is 5\,000 years into 
this phase, in panel (b) 10\,000 years, in panel (c) 20\,000 years and 
in panel (d) 30\,000 years. The position of the central star is marked 
by an asterisk. The colour scaling is logarithmic and the density is 
scaled in units of \density\ where $2.5\times10^6$ is equivalent to 
$n_{\rm H}$ = 100 cm$^{-3}$.}
\label{pn2}
\end{center}
\end{figure*}

In Figure \ref{pn2} we show the results of case B during the 
post--AGB phase. We do not show the AGB phase as the structure formed 
during this phase is clear in panel (a), where the PN has been expanding 
for 5000 years into the AGB wind bubble behind the bow shock. The central star 
is now moving at 50 \kms, an average \vism\
for a PN-forming star. A bow shock has formed ahead of the central star
during the AGB phase and stabilised at the point of ram pressure balance
0.25 pc ahead of the central star. This is a much smaller structure than
in case A due to the increased ram pressure of the ISM.
Further, the bow shock in this case is not smooth;
instead there are indications of turbulent motion which can be seen
originating in the AGB phase as instabilities at the head of the bow 
shock and moving back down the tail over tens of thousands of years.
Note that the region between the forward shock and contact discontinuity 
is compressed upstream compared to the previous case. An investigation 
of the temperatures in this region shows that the material is far hotter 
at $\sim30\,000$ K than in the previous case, in agreement with strong 
shock predictions, and therefore stronger radiative cooling can be inferred 
to be the cause of the forward shock compression. The PN is still inside
the bubble of undisturbed AGB wind material and as yet unaffected by
the ISM.

In panel (b), 10\,000 years into the PN phase, the PN has reached the
second stage of PN--ISM interaction whereby the PN still appears circular but
is now rebrightened at the interaction with the AGB wind bow shock. The PN is also rapidly
entering a third stage where the geometric centre of the nebula, defined as
the centre of the circular/elliptical shape on the sky, is moving away from
the central star.

After 20\,000 years of PN evolution, the PN--ISM interaction has distorted the
circular shape of the nebula as shown in panel (c). The central star appears displaced
by a quarter of the diameter upstream of the geometric centre.  The highest
density and temperature regions have now moved around the bow shock to the
point where the tail stretches away from the PN shell. These regions can be
interpreted as areas where the nebular shell and ISM have driven the AGB wind
material away from the head of the bow shock towards the tails.  It is worth
noting that at this time estimates of the nebular age via its observational
dynamics (i.e. gas expansion velocity) would underestimate its true age.

In panel (d), 30\,000 years into the PN phase, the PN has further
departed from circular symmetry on the sky. The regions of highest density and
temperature are towards the head
of the fast wind bow shock which is now forming ahead of the star against the
oncoming ISM. The nebular shell is still progressing downstream though the
undisturbed AGB wind bubble and causing the geometric centre of the nebula to
deviate downstream. This stage of significant deceleration of the upstream
shell was considered by \cite{soker91}.

At this typical \vism, PN--ISM interaction has strongly affected
the evolution of the PN. Rebrightening via interaction has occured at a much
younger age of only a few thousand years and the shape of the PN has deviated
strongly from circularity on the sky.  One sided brightening and/or deviations
from circular symmetry combined with an off-geometric-centre central star are
clearly strong indicators of PN--ISM interaction. Further, these indicators are
not limited to ancient or diffuse PNe. Such an interaction at this average
speed would indicate many PNe should display some characteristics of ISM
interaction during their later life.

\subsection{Case C}

\begin{figure*}
\begin{center}
\includegraphics[angle=0,width=14cm]{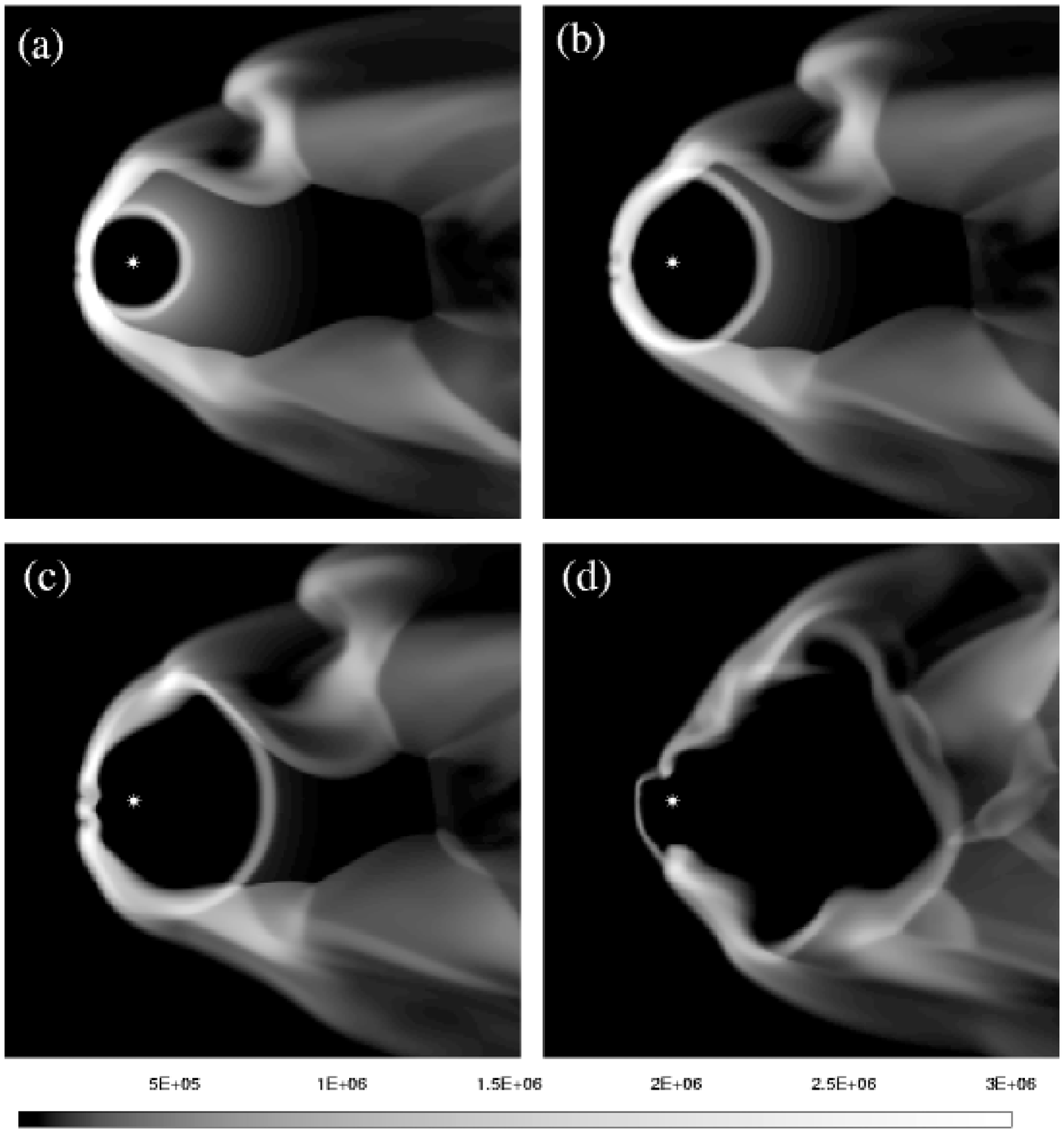}
\caption{The results of case C: the panels show the gas density during
the post--AGB phase. In panel (a), the simulation is 5000 years into this 
phase, in panel (b) 10\,000 years, in panel (c) 15\,000 years and in 
panel (d) 30\,000 years. The position of the central star is marked by 
an asterisk. The colour scaling is logarithmic and the density is 
scaled in units of \density\ where $2.5\times10^6$ is equivalent to 
$n_{\rm H}$ = 100 cm$^{-3}$.}
\label{pn3}
\end{center}
\end{figure*}

In Figure \ref{pn3}, we show the results of case C during the post--AGB
phase. In panel (a), 5000 years into this phase, the PN already has passed
through the first stage of interaction and the large higher density region on
the upstream side of the PN indicates it is now some way into the second
stage.  The stabilisation of the bow shock a shorter distance of 0.18-pc
upstream is responsible for this earlier interaction. Note that with
\vism\ = 75 \kms\ the bow shock is now seriously disrupted and
instabilities noted in the previous case are now forming cool, high-density
spiralling vortices moving down the tail. The origin of these vortices as
vortex-shedding instabilities at the head of the bow shock and their effect on
the local ISM are discussed elsewhere \citep{wareing07}. We find the vortices
will affect almost all bow shock structures and PNe will interact with them at
some point during their evolution. A higher speed case carried out by
VGM also revealed similar structures forming in the AGB wind bow shock
\citep{szentgyorgyi03}.

In panel (b), the upstream portion of the nebular shell has been significantly
decelerated causing the geometric centre of the nebula to shift downstream.
At this stage, 10\,000 years into the post--AGB phase, the regions of highest
density and temperature have again moved round the bow shock
and the PN is in the third stage of PN--ISM interaction.

In the third panel, 15\,000 years into the post--AGB phase, the object is still
recognisable as a PN, although the regions of highest density 
are in the tail of the bow shock. Finally, after 30\,000 years of post--AGB
evolution, the densest regions of the PN are indicative of the structure of
the tail of the bow shock, including the effect of the vortices moving
downstream.  The thin fast wind bow shock is clear ahead of the PN-forming
star, although the star is no longer anywhere near the centre of the
object. This would be an ancient and probably very faint PN, but the PN--ISM
interaction would still cause many parts of it to rebrighten.

In summary, the PN--ISM interaction is becoming more important at higher
\vism:  the bow shock is closer to the star causing
the PN to move through the stages of interaction more quickly.
PN--ISM interaction would be apparent at a much earlier age even at this
relatively average \vism\ of 75 \kms.

\subsection{Case D}

\begin{figure*}
\begin{center}
\includegraphics[angle=0,width=14cm]{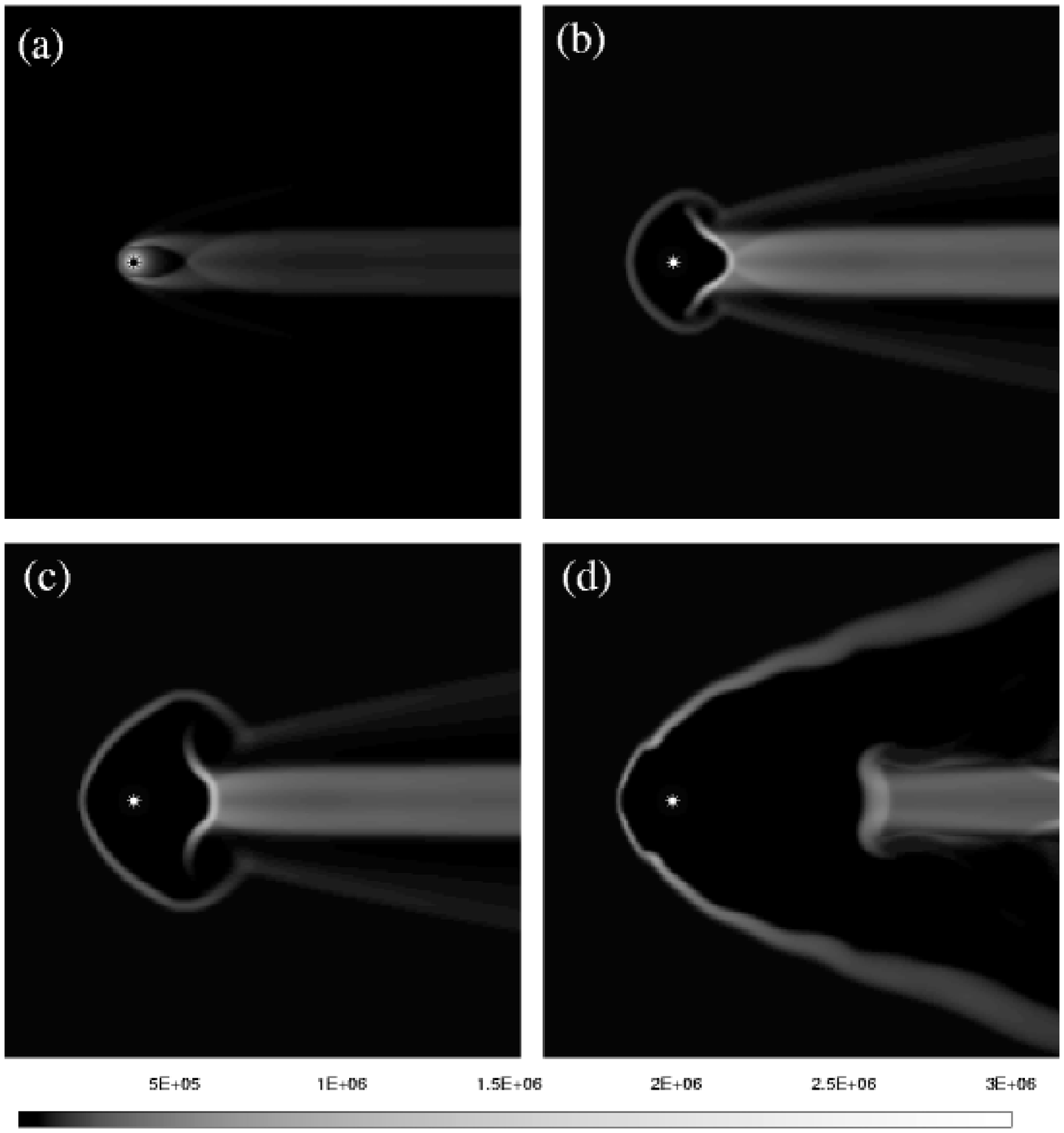}
\caption{The results of case D: the panels show the gas density during
the post--AGB phase. In panel (a), the simulation at the end of the AGB
phase, in panel (b) 1000 years
into the post--AGB phase, in panel (c) 2000 years and in panel (d) 10\,000 years. 
The position of the central star is marked by an asterisk. The colour
scaling is logarithmic and the density is scaled in units of \density\ 
where $2.5\times10^6$ is equivalent to $n_{\rm H}$ = 100 cm$^{-3}$.}
\label{pn4}
\end{center}
\end{figure*}

In Figure \ref{pn4} we show the results of case D, where the central
star has \vism\ = 100 \kms. A narrow confined bow shock has formed ahead of 
the star, with a tail stretching far downstream, as shown in panel (a), by
the end of the AGB phase. This central star could be considered much like a
speeding bullet and the high \vism\ and high ISM density combined
with the low mass-loss rate of the slow wind has resulted in a much more
confined bow shock.  The ISM has exerted a naturally stronger shaping
influence due to greater ram pressure. Any instabilities at the head of the
bow shock are either smoothed out by the oncoming ISM or flow out of the
simulation before they have time to fully form.

In panel (b), 1000 years into the PN phase, the PN has very rapidly interacted
with the bow shock and formed a fast wind bow shock further ahead of the
central star, which can also be understood in terms of a ram pressure balance.
Interestingly, the highest densities are in the region of the nebular shell
which is moving downstream. In contrast to earlier cases, this portion
of the shell is now closest to the star. It has rapidly moved through the
small bubble of undisturbed AGB wind material and is now interacting with the
more dense shocked AGB wind material in the tail. In panel (c), after 2000 years,
this effect is even more apparent. Over the course of the next few thousand
years, the tail of the wide fast wind bow shock grows downstream and the
remaining part of the nebular shell continues to move away from the central
star through the tail of older shocked AGB wind material.

The PN formed in this case is very different to many others and characteristic
of small and confined bow shocks throughout our simulations. How observable
the fast wind bow shock would be remains an open question as whilst it is hot,
it is also of low density. The PN in this case has rapidly moved through the first three
stages of PN--ISM interaction. We interpret the stage where the fast wind bow
shock is forming and the remains of the nebular shell are fading whilst moving
downstream as the fourth and final stage of PN--ISM interaction. During this
stage, it is difficult to recognize the object, which would still be young
enough to be observable, as a PN.

\subsection{Case E}

\begin{figure*}
\begin{center}
\includegraphics[angle=0,width=14cm]{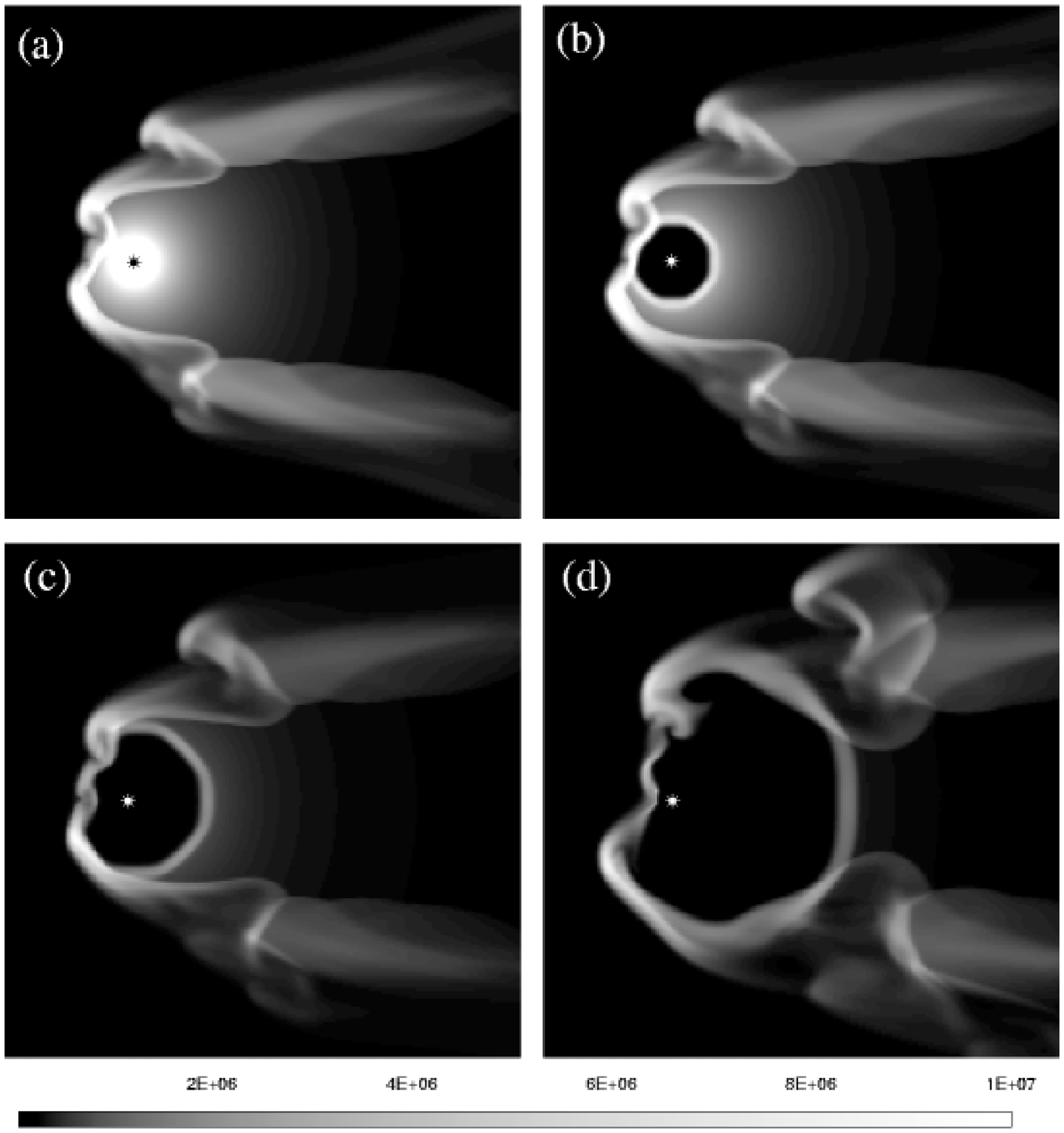}
\caption{The results of case E: the panels show the gas density during
the post--AGB phase. In panel (a), the simulation at the end of the AGB
phase and the beginning of the post--AGB phase, in panel (b) 2000 years
into the post--AGB phase, in panel (c) 4000 years and in panel (d) 10\,000 
years. The position of the central star is marked by an asterisk. The
colour scaling is logarithmic and the density is scaled in units of 
\density\ where $5\times10^6$ is equivalent to $n_{\rm H}$ = 200 cm$^{-3}$.}
\label{pn5}
\end{center}
\end{figure*}

The final case under discussion in this paper is case E, shown in Figure
\ref{pn5}. In this case the central star has a high \vism\ of 125
\kms, a high ISM density and a high slow wind mass-loss rate.  The bow shock
structure is comparatively large because of the high mass-loss rate.
Instabilities in the bow shock have caused multiple vortices to be shed
downstream with an ensuing complex bow shock structure, as shown in panel (a) at
the end of the AGB phase. The PN--ISM interaction is almost immediately
apparent and after 2000 years, as shown in panel (b), the PN is mid-way through the second
stage of interaction. The deceleration of the PN shell caused by the bow
shock will soon cause the geometric centre to shift downstream and the PN to
enter the third stage of interaction.  Observing a PN with a high-velocity
central star and a faint completion of the nebular ring moving downstream
would indicate a young rather than old PN, even though the PN--ISM interaction
is strongly apparent.  In panel (c), 4000 years into the post--AGB phase, the 
densest regions of the PN have moved away from the head of the bow
shock. The structure of the bow shock at the end of the AGB phase is having a
very strong effect on the appearance of the PN -- particularly the fact that
the bow shock had fallen back towards the central star having just shed a
vortex-causing instability downstream. In later stages (e.g. after 10\,000
years shown in panel (d)) the vortices have become even more important for 
the evolution of the PN and are the regions of highest density.
The object will eventually fade and look even less like a PN;
faint objects serendipitiously discovered as part of Galactic surveys,
e.g. IPHAS \citep{drew05}, may closely resemble this PN whilst not be
classified as such.  Eventually, after the central star has turned off and
there is no longer a wind supporting the bow shock, the nebula would be
rapidly blown  downstream and the central star would move outside its
nebula.

\section{Discussion}

\subsection{The four stages of interaction}

\begin{figure}
\begin{center}
\includegraphics[angle=0,width=7cm]{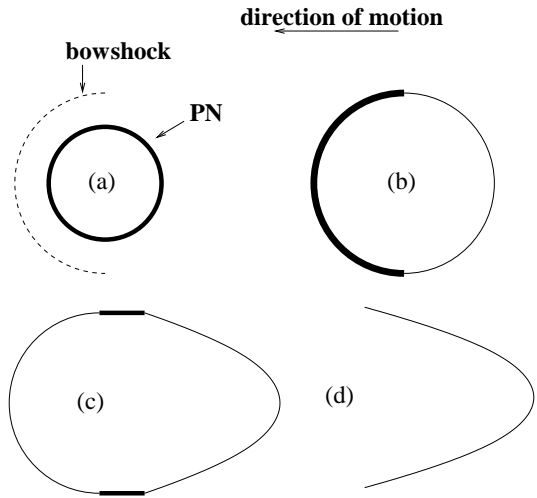}
\caption[The four stages of PN--ISM interaction]{A simple illustration of the 
appearance of a PN during the four stages of PN--ISM interaction discussed in
section 5.1. The direction of motion is to the left and thicker lines indicate 
the brightest regions. Panel (a) illustrates stage WZO 1, (b) stage WZO 2), (c) 
stage WZO 3 and (d) stage WZO 4. The position of letters (a), (b), (c) and (d) 
indicate the position of the central star at each stage.}
\label{pnstages}
\end{center}
\end{figure}

How a PN will interact with the ISM is set during the AGB
phase and VGM was the first to highlight this important
point. We have identified four stages of interaction which we call WZO 1--4.
There is a phase of evolution where the PN is expanding within the
bubble of undisturbed AGB wind material and this we define as the first stage
of PN--ISM interaction, WZO 1. During this stage, a PN would be unaffected by the ISM
interaction which is radially further from the central star. We suggest it
would be possible during this stage to observe a faint arc around the PN,
which would be the AGB wind bow shock. In the case of a slow-moving star with
a large bow shock, this stage can last for the entire lifetime of the PN and
it is unlikely a PN--ISM interaction will ever be observed. However, if the
central star is moving even at average speed, the PN--ISM interaction can
become rapidly apparent, in some of our simulations after only a thousand years. We
show this stage in Figure \ref{pnstages}(a).

The characteristic of stage 1 is a shell of swept-up ISM, up to a few pc
away. This has been called a 'wall' \citep{zijlstra02}. On sky images, this
may also show up an area of reduced emission around the PN, ending at the
wall. Evidence for such cavities has been presented by \citet{evans02} and
\citet{weinberger03}.

Stage WZO 2 is entered when the PN has expanded far enough to interact with
the bow shock formed during the AGB phase of evolution. As the PN shock merges
with the AGB wind bow shock, driving another shock through it, the density and
temperature of the material increase accordingly and strengthen the emission
from this area.
This is shown in Figure \ref{pnstages}(b).  If \vism\ is predominantly in the
plane of the sky, we would observe part of the nebular shell brighter than the
rest.  If, however, \vism\ is almost all along the line of
sight to the PN, emision from the whole ring structure would strengthen and it would be difficult
to identify this PN as undergoing a PN--ISM interaction until later in its
evolution when distortions of the PN shell may reveal its true nature.  This
stage is relatively short lived and can be as short as a thousand years or so
in our simulations with the largest \vism. During this stage the whole
PN continues to appear circular on the sky, but as the central star moves away
from the geometric centre of the nebula, caused by the deceleration of the
nebular shell in the direction of motion, the PN enters the third stage of
PN--ISM interaction.

The third stage of interaction, WZO 3, is defined by the geometric centre of the
nebula moving downstream away from the central star as shown in
Figure \ref{pnstages}(c). The shift of the geometric centre due to the
deceleration of the PN shell in the direction of motion is guaranteed to occur
and provides a measurable effect of this interaction.  During this stage, and
beginning during the second stage, it is difficult to estimate the age of the
PN from its apparent diameter on the sky. The AGB wind--ISM interaction has
coccooned the PN inside the AGB wind bow shock and this considerably inhibits the
expansion of the PN shell during stages two and three. Identification of the
arc of nebular shell moving downstream yet still inside the AGB wind bubble and the
central star would provide an estimate of the radius of the PN if it had not 
been inhibited by the PN--ISM interaction. This radial distance could be used
to dynamically estimate the age of the PN. Interestingly, our simulations suggest that a
secondary effect during this stage is that the regions of highest density and
temperature move away from the head of the bow shock. This can be understood
as the oncoming ISM sweeping the shocked PN material from the second stage
towards the tail of the nebula. This third stage can be open-ended
and only in the higher \vism\ cases is the PN affected enough by
the interaction to become completely unfamiliar and enter the fourth stage of
interaction before it fades away. Even in the highest velocity cases, the
third stage lasts at least 10\,000 years. It is possible that in the most
extreme cases, the PN shell will continue to appear circular as the nebula is
swept downstream of the central star. Sh\,2-68 is an example of a nebula where
the central star is thought to have deserted its PN \citep{kerber02}.

In the fourth and final stage of PN--ISM interaction, WZO 4, the PN no longer appears
circular. The fast wind has formed a bow shock ahead of the star and the
little remaining AGB wind material in the vicinity of the star is being swept
downstream with turbulent areas of high density and temperature as shown in
Figure \ref{pnstages}(d). At this time, the observable structure 
may not be identified as a PN. Further, the central star appears to have
long since left these regions. Many objects such as this may exist and
are probably not classified as PNe, affecting estimates of the Galactic 
distribution of PNe. Deep surveys in particular will be abe to
uncover these objects.

Our four stages of the interaction are summarised in Table \ref{summary}. 
The first two stages are similar to the stages of evolution
suggested by \cite{borkowski90} from their observations of ancient PNe,
although we find they can occur earlier in the PN evolution than
considered in that work, due to the pre-shaping during the AGB.

\begin{table*} 
\caption[The four stages of PN--ISM interaction]{The four stages of PN--ISM
interaction and their observable effects, as discussed in section 5.1}
\label{summary}
\begin{center}
\begin{tabular}{cc} 
{\it stage} & {\it observable effects} \\
\hline
WZO 1 & PN as yet unaffected; faint bow shock may be observable.\\
WZO 2 & Brightening of PN shell in direction of motion.\\
WZO 3 & Geometric centre shifts away from central star,\\
WZO 4 & PN completely disrupted, central star is outside the PN.\\
\hline
\end{tabular}
\end{center}
\end{table*}

Central stars of PNe which show evidence of interaction should have a 
proper motion consistent with the observed distortions across the plane of 
the sky if the nebula is close enough and the central star is moving fast 
enough in the right direction to have an appreciable angular 
motion over time. \cite{borkowski90} performed an investigation of PNe 
with known large angular motion and revealed many show signs of being
in WZO Stage 2. 

The largest value of proper motion of a central star measured via ground-based 
observations is $53.2 \pm 5.5$ mas yr$^{-1}$. This is the proper motion of the 
central star of Sh\,2-68 \cite{kerber02}. The measurement has provided direct 
confirmation of the process of the central star being displaced from the 
geometric centre of the nebula.

\subsection{Missing mass}

The interaction with the ISM considerably alters the amount of mass within the
observed nebula: the ram-pressure stripping of material downstream during the
AGB phase removes mass from the circumstellar region.
Our simulations show that up to 90 per cent of the mass ejected from the star
during the AGB phase can be left downstream forming the tail behind the
nebula. This effect may provide a solution to the missing mass problem in PNe
whereby only a small fraction of the mass ejected during the AGB phase is
observationally inferred to be present during the post--AGB phase.  Stellar
evolution calculations predict that stars with initial masses in the range of
1--5 M$_\odot$ will end as PN nuclei with masses around 0.6 M$_\odot$. Most of
the mass is lost on the AGB phase and should be easily observable as ionized
mass during the PN stage. However, observations of Galactic PNe reveal on
average only 0.2 M$_\odot$ of ionized gas. As the interaction progresses, the
mass in the PN shell is increased during merger with the AGB wind bow shock,
but this effect is minimal when compared to the mass left downstream. At
higher speeds, the stripping effect is greater and more mass is lost
downstream. Our simulations clearly support VGM's conclusion that PN--ISM
interaction at low speeds can provide an explanation of the missing mass
phenomenon. Further, we show that this effect is even more pronounced at high
speed.

Recent observations of the Mira system containg the AGB star Mira A have
revealed a comet-like tail of material stretching 4 pc North away from 
the system (assuming a distance of 107 pc) and an arc-like structure 
in the South \citep{martin07}. Martin et al go on to comment that the 
space velocity of 130 \kms\ is in a direction consistent with the 
comet-like tail being a ram-pressure-stripped tail of material behind 
a bow shock ahead of the system, thereby confirming our postulation of 
tails of ram-pressure-stripped material behind AGB stars.

\subsection{Limitations of the model}

In reality, it is reasonable to expect the AGB wind to show an increasing
mass-loss rate with time, whilst the post--AGB wind may increase in velocity
over time. We have shown that our current assumptions are sufficient to
reproduce basic nebular structure and leave detailed temporal modelling to a
future publication. 

In our models, we have not considered the intrinsic
structure of the PN; as initially discussed, PN display many asymmetric shapes
which cannot be explained in the context of this model.  The inclusion of
time-variant AGB and post--AGB wind parameters, asymmetric winds and/or
magnetic fields may address this, but this is beyond this publication.

In contrast to the studies of \cite{dgani94} and \cite{dgani98}, we have
observed no fragmentation in our model above 100 \kms.  In an effort to
understand whether grid resolution affects fragmentation, we have run a high
resolution simulation on the head of the bow shock with effectively 8 times
higher resolution and found no evidence of fragmentation.  This disagreement
could be attributed to the simulations being of too low resolution to allow
instabilities to fragment the bow shock in such a way as theoretically
predicted. \cite{soker91} did not consider evolution on the AGB which alters
the structures considerably.  Our simulations have not included the effect of
magnetic field, neither as a scalar pressure nor a vector field. Magnetic
fields, if inclined to the direction of relative motion, break the cylindrical
symmetry of the interaction process which can affect these instabilities and
lead to the formation of elongated structures. \cite{huggins05} extends the
suggestion that filaments observed in PNe may be signatures of an underlying
magnetic field. We have demonstrated that magnetic fields are not necessary to
explain asymmetries observed in slow-moving PNe.  It is possible that the
inclusion of a gravitational field in the hydrodynamic method may fragment the
bow shocks but this is also beyond the scope of this work. Further, an
inhomogeneous ISM could alter the structure of the bow shock.

In our simulations, we see instabilities forming at the head of the AGB wind
bow shock causing vortex shedding downstream. When the PN shell expands far
enough to interact with these vortices, typically during stage 3 or 4, it is
affected and the PN shell would be distorted and brightened accordingly. We
have discussed the importance of these vortices for the local ISM elsewhere
\citep{wareing07}. Inhomogeneities in the ISM could be expected to seed more
instabilities for vortices.

Finally, we note that observability of such objects is dependent on the extent
of the photoionisation and on excited ionisation lines. 

\section{Comparisons to observed PNe}

\subsection{Relevance for Galactic Disk PNe}

We have briefly examined the IAC Morphological Catalog of Northern
Galactic Planetary Nebulae \citep{manchado96} and list in Table \ref{nebulae} the PNe which
we suggest are interacting with the ISM, the stage of their PN--ISM interaction and 
their brief interaction characteristics.

\begin{table*} 
\caption{Nebulae displaying characteristics of ISM interaction from the
IAC Morpohological Catalog of Northern Galactic Planetary nebulae
\protect\citep{manchado96} discussed in Section 6.1}
\label{nebulae}
\begin{center}
\begin{tabular}{ccl} \hline
Name & Stage of & Charecteristics \\
 & interaction & \\
\hline
A\,13 & WZO 2 & bow shock/PN interaction to the West \\
A\,16 & WZO 2 & bow shock/PN interaction to the South-East \\
A\,52 & WZO 2 & bow shock/PN interaction to the North-East \\
A\,58 & WZO 2 & brightened on the Western side; may be a bow shock interaction \\
A\,59 & WZO 2 & bow shock/PN interaction to the North \\
A\,86 & WZO 2 & bow shock/PN interaction to the North-East \\
Ba\,1 & WZO 1 & faint bow shock structure outside PN to the North-West \\
DeHt\,2 & WZO 1 & faint bow shock structure outside PN to the North \\
DeHt\,4 & WZO 2 & bow shock/PN interaction to the West \\
EGB\,4 & WZO 3/4 & narrow, confined (fast wind?) bow shock to the South \\
He\,2-428 & WZO 2 & bow shock/PN interaction to the South \\
IC\,4593 & WZO 1 & faint bow shock structure outside PN to the North-West \\
Jn\,1 & WZO 3 & bow shock/PN interaction with emission shifted downstream \\
K\,2-2 & WZO 3/4 & structure may be a bow shock remnant \\
K\,4-5 & WZO 3/4 & structure may be remnant of a bow shock blown downstream \\
M\,2-2 & WZO 2 & bow shock/PN interaction to the South \\
M\,2-40 & WZO 1 & faint bow shock structure outside PN to the East \\
M\,2-44 & WZO 1 & faint bow shock structure outside PN to the South-West \\
NGC\,6765 & WZO 1 & faint bow shock structure outside PN to the West \\
NGC\,6853 & WZO 1 & faint bow shock structure outside PN to the North-West \\
NGC\,6891 & WZO 1 & faint bow shock structure outside PN to the South-West \\
S\,22 & WZO 2 & bow shock/PN interaction to the South-East \\
Sd\,1 & WZO 2 & bow shock/PN interaction to the North \\
\hline
\end{tabular}
\end{center}
\end{table*}

Of the approximately 130 well-resolved nebulae in the catalogue, if we assume
a random distribution of angles between \vism\ and the line of sight then
approximately 15--20 per cent of the nebulae will have their motion
predominantly in the plane of the sky. PN--ISM interaction is clear
in cases where \vism\ is in the plane of the sky and our brief inspection
of the catalogue has revealed 20 per cent show characteristics of PN--ISM 
interaction. The correlation between these statistics supports our 
finding that interaction is common, particularly amongst large 
and/or evolved PNe. A further study of our selected PNe may reveal angular 
motions of the central stars, which should be in the direction of the ISM 
interaction. 

Considering the list of PNe given by \cite{borkowski90}, we find that all
of them are in the WZO 1 or WZO 2 stages of PN--ISM interaction.

\subsection{Relevance for Galactic Bulge PNe}

Galactic Bulge PNe have central stars with higher typical \vism. 
Our simulations have shown that simulations with higher typical \vism\ 
are generally smaller and show signs of strong interaction at an earlier age,
thus we would expect Galactic Bulge PNe to have such characteristics.

\subsection{Relevance for PNe in globular clusters}

Three PNe are known in globular clusters. The PN in M15 (K648)
shows a strongly edge-brightened structure, reminiscent of a WZO 2 interaction
 \citep{alves00}. The PN in M22 is known to be peculiar, strong in [O\,III]
but absent in H$\alpha$: it is worth investigating the possible effect of
shock-excitation on its spectrum. Its structure shows  a clear bow shock
\citep{borkowski93}, with the star close to the parabolic arc. This can be
identified with a type WZO 2 interaction. The tail is not seen.

\begin{figure}
\begin{center}
\includegraphics[angle=0,width=7cm]{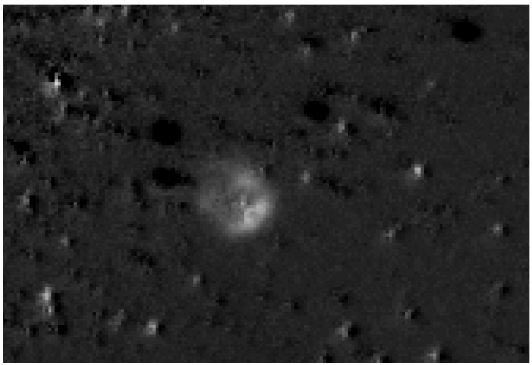}
\caption{VLT FORS1 image in [O\,III] of the PN in NGC\,6441. Continuum
has been subtracted. The edge brightening is indicative a PN--ISM shock.}
\label{ngcpn}
\end{center}
\end{figure}

The final confirmed globular cluster PN \citep{jacoby97} is in the cluster NGC
6441. No previous high-resolution images have been published. We obtained an
image taken with the VLT in [O\,III], background subtracted, courtesy of
M. van Haas and L. Kaper. The image is shown in Fig. \ref{ngcpn}: it shows a
clear edge brightening, with the brightest region possibly split from the head
of the bow shock. This is indicative for an early type WZO 3 interaction. We
conclude that all three globular cluster PNe are dominated by ISM interaction.

As globular clusters have been stripped of their interstellar gas and are
moving at high velocities (up to 200 \kms) through the Galactic halo, we would
expect to see severe effects for these short-lived objects including rapid
brightening via interaction and destruction.

\section{Conclusions}

Observationally, the presence of asymmetries in the haloes of PNe has been
found to be a relatively common feature \citep{tweedy96, guerrero98}. These
asymmetries can be partially if not wholly attributed to interaction with the
ISM. The high rate of asymmetries can now be explained when the evolution
through the AGB phase of the central star is considered. The simulations in
this work show that interaction is present at all evolutionary stages for all
\vism.

Our comprehensive simulations have reinforced the result of VGM that PN--ISM
interaction is set during the AGB phase and affects central stars with
typical \vism, not just those with extreme velocities. We have
developed a model which results in four stages of ISM interaction and 
inspections of PNe support this four stage interpretation.

PNe have been found to evolve more quickly through these stages the faster they 
move through the ISM. They have also been found to be smaller for greater \vism. 
Much of their mass ejected during the AGB phase has been stripped 
downstream into the tail, providing a possible explanation of the missing mass problem 
observed in PNe, with evidence for such a tail found by \cite{martin07}. 
The average speed results of the model appear very similar to 
the recently discovered bow shock surrounding the Dumbbell Nebula 
which has an average \vism\ \citep{meaburn05} further supporting 
our model.

Our conclusions are in agreement with VGM using similar models.
We have extended the study of the PN--ISM interaction including the AGB phase
to three dimensions and higher \vism\ where they were previously
limited to two dimensions and low \vism.

PNe which show signs of ISM interaction are not necessarily ancient, nor
require a high \vism\ or magnetic fields. Interaction can become
apparent at a young age via several methods and central stars with an average
\vism\ can show evidence of interaction in their nebulae, as commonly observed, 
and be displaced from the geometric centre of their nebulae. None of the
simulations have required a magnetic field to produce commonly observed
effects, although the fragmentation of the bow shock in PNe as Sh~2-188 awaits
a fuller treatment of this interaction at higher resolutions involving
inhomogeneous ISM and gravitational and magnetic field modelling to reasonably
test the fragmentation theories of interacting PNe.

\section*{Acknowledgments}

This work was carried out as part of CJW's STFC-funded PhD project at Jodrell Bank
under the supervision of TOB and as part of STFC rolling grant-funded
post-doctoral research at the University of Manchester.
The numerical computations were carried out using the Jodrell
Bank Observatory COBRA supercomputer.

\begin{appendix}
\section{Appendix 1}

In figures 1--8, we show slices of the simulation domain through the
position of the central star, parallel to the direction of motion at
the end of the AGB phase of evolution.
Each figure shows the set of simulations at a particular \vism.
In the accompanying tables 1--8, we list the parameter values for each
simulation.

\pagebreak

\begin{figure*}
\begin{center}
\includegraphics[angle=0,width=16cm]{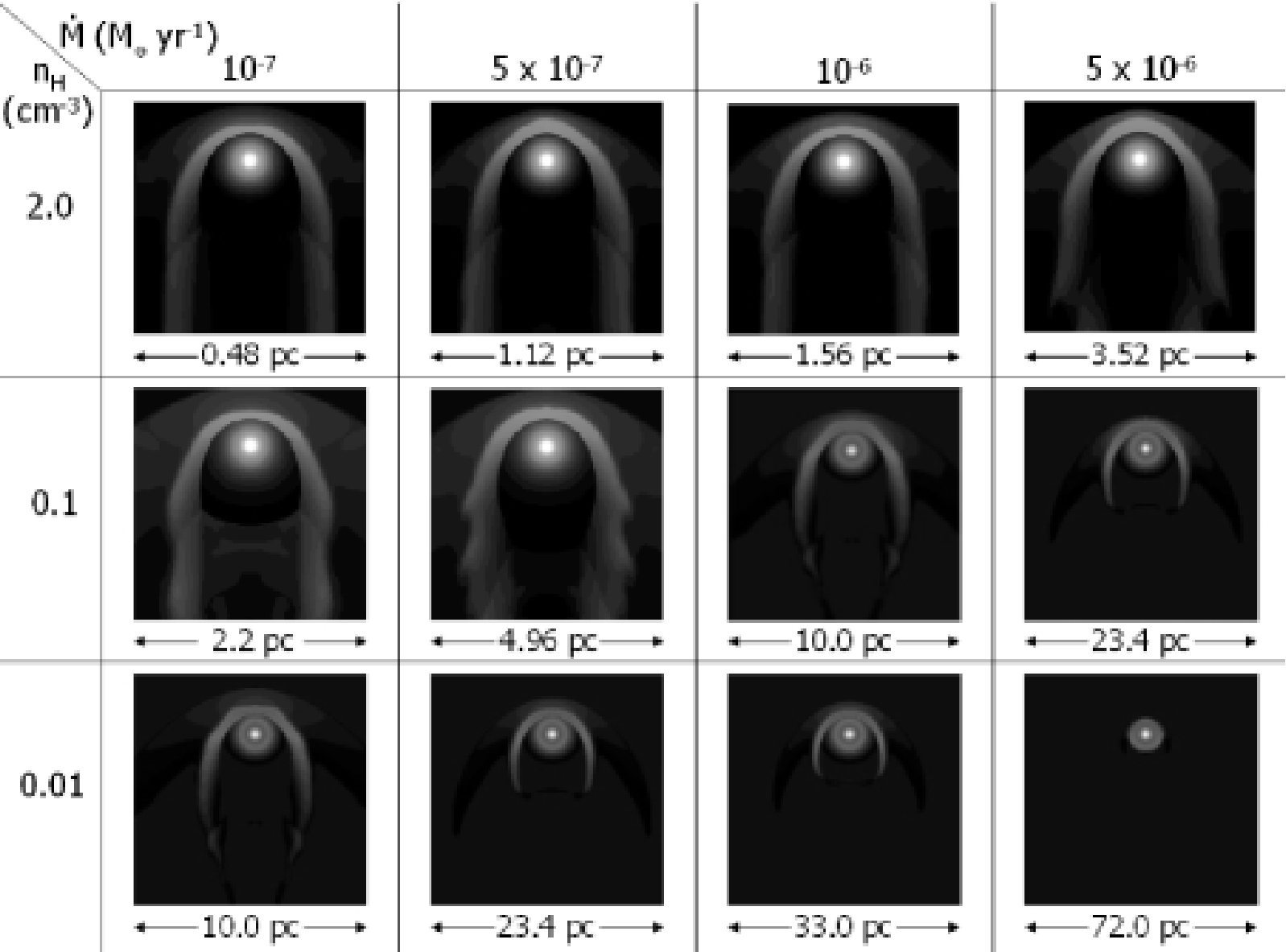}
\caption{Snapshots at the end of the AGB phase of evolution for 
\vism\ = 25 \kms. mass-loss rate increases from left to
right, ISM density decreases from top to bottom. Full details of each 
simulation can be found in table \ref{apptable-25}.}
\label{app-25}
\end{center}
\end{figure*}

\nopagebreak

\begin{table*}
\caption{Input parameters for the PN--ISM simulations in figure \ref{app-25}: 
column a) gives the mass-loss rate in the slow wind; column b) 
the density of the surronding ISM in $n_{\rm H}\,\rm cm^{-3}$;
column c) the relative velocity of the central star; 
column d) the grid dimension along one side.}
\label{apptable-25}
\begin{center}
\begin{tabular}{ccccc}
\hline
a) & b) & c) & d) &\\
$\dot{M}_{\rm sw}$ & $\rho_{\rm ISM}$ & \vism & Grid &\\
(M$_{\odot}$\,yr$^{-1}$) & $n_{\rm H}$(cm$^{-3}$) & (\kms) & (pc) & \\
\hline
$10^{-7}$  & 2 & 25 & 0.48 & \\
$5\times10^{-7}$  & 2 & 25 & 1.12 & \\
$10^{-6}$  & 2 & 25 & 1.56 & \\
$5\times10^{-6}$  & 2 & 25 & 3.52 & \\
$10^{-7}$  & 0.1 & 25 & 2.2 & \\
$5\times10^{-7}$  & 0.1 & 25 & 4.96 & \\
$10^{-6}$  & 0.1 & 25 & 10.0 \\
$5\times10^{-6}$  & 0.1 & 25 & 23.4 & \\
$10^{-7}$  & 0.01 & 25 & 10.0 & \\
$5\times10^{-7}$  & 0.01 & 25 & 23.4 & case A\\
$10^{-6}$  & 0.01 & 25 & 33.0 & \\
$5\times10^{-6}$  & 0.01 & 25 & 72.0 & \\
\hline
\end{tabular}
\end{center}
\end{table*}

\clearpage

\begin{figure*}
\begin{center}
\includegraphics[angle=0,width=16cm]{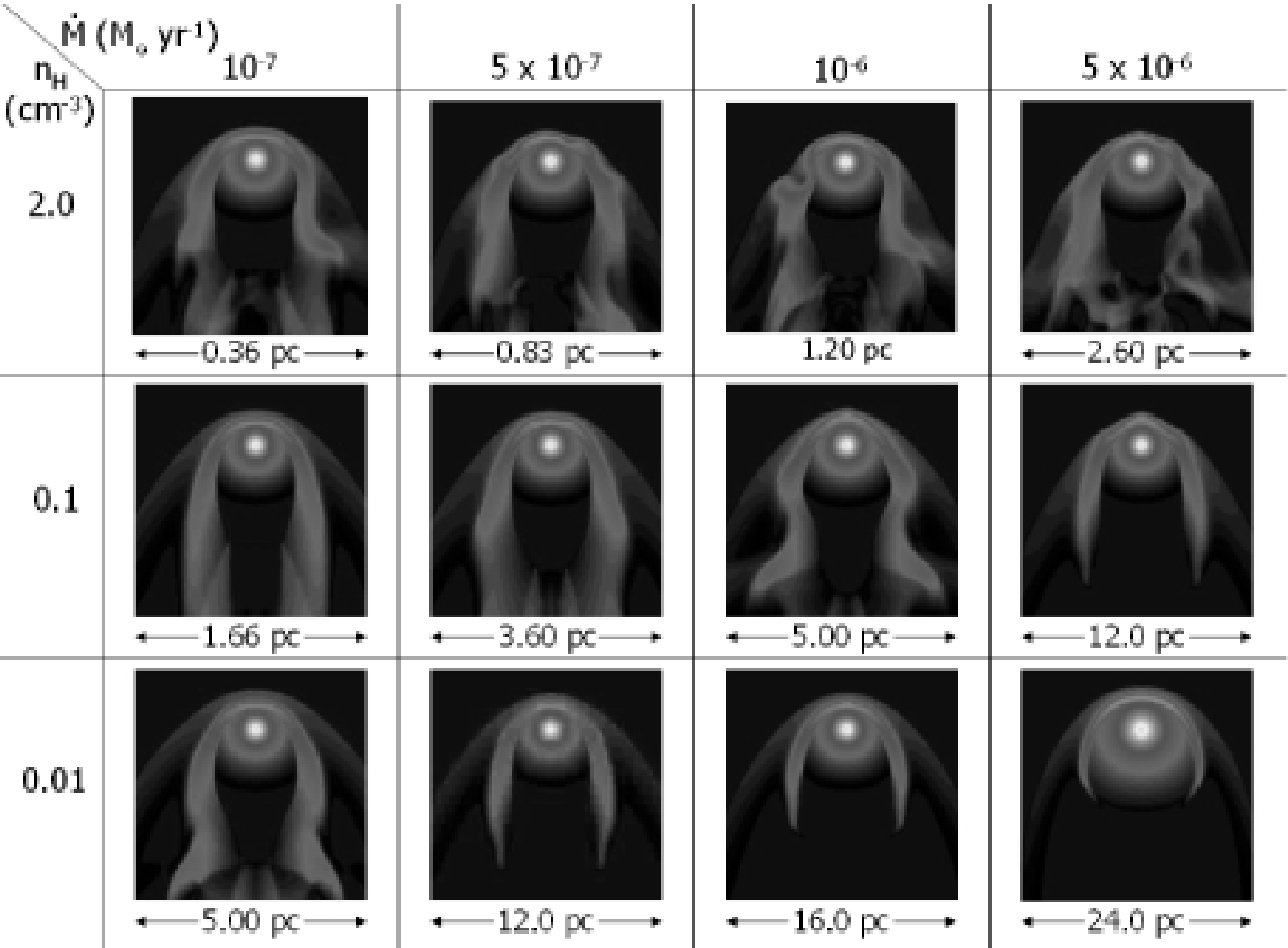}
\caption{Snapshots at the end of the AGB phase of evolution for
\vism\ = 50 \kms. mass-loss rate increases from left to
right, ISM density decreases from top to bottom. Full details of each 
simulation can be found in table \ref{apptable-50}.}
\label{app-50}
\end{center}
\end{figure*}

\begin{table*}
\caption{Input parameters for the PN--ISM simulations in figure \ref{app-50}: 
column a) gives the mass-loss rate in the slow wind; column b) 
the density of the surronding ISM in $n_{\rm H}\,\rm cm^{-3}$; 
column c) the relative velocity of the central star; 
column d) the grid dimension along one side.}
\label{apptable-50}
\begin{center}
\begin{tabular}{ccccc}
\hline
a) & b) & c) & d)& \\
$\dot{M}_{\rm sw}$ & $\rho_{\rm ISM}$ & \vism & Grid & \\
(M$_{\odot}$\,yr$^{-1}$) & $n_{\rm H}$(cm$^{-3}$) & (\kms) & (pc)   & \\
\hline
$10^{-7}$  & 2 & 50 & 0.36 & \\
$5\times10^{-7}$  & 2 & 50 & 0.83 & \\
$10^{-6}$  & 2 & 50 & 1.20 & \\
$5\times10^{-6}$  & 2 & 50 & 2.60 & case B\\
$10^{-7}$  & 0.1 & 50 & 1.66 & \\
$5\times10^{-7}$  & 0.1 & 50 & 3.60 & \\
$10^{-6}$  & 0.1 & 50 & 5.0 & \\
$5\times10^{-6}$  & 0.1 & 50 & 12.0 & \\
$10^{-7}$  & 0.01 & 50 & 5.0 & \\
$5\times10^{-7}$  & 0.01 & 50 & 12.0 & \\
$10^{-6}$  & 0.01 & 50 & 16.0 & \\
$5\times10^{-6}$  & 0.01 & 50 & 24.0 & \\
\hline
\end{tabular}
\end{center}
\end{table*}

\begin{figure*}
\begin{center}
\includegraphics[angle=0,width=16cm]{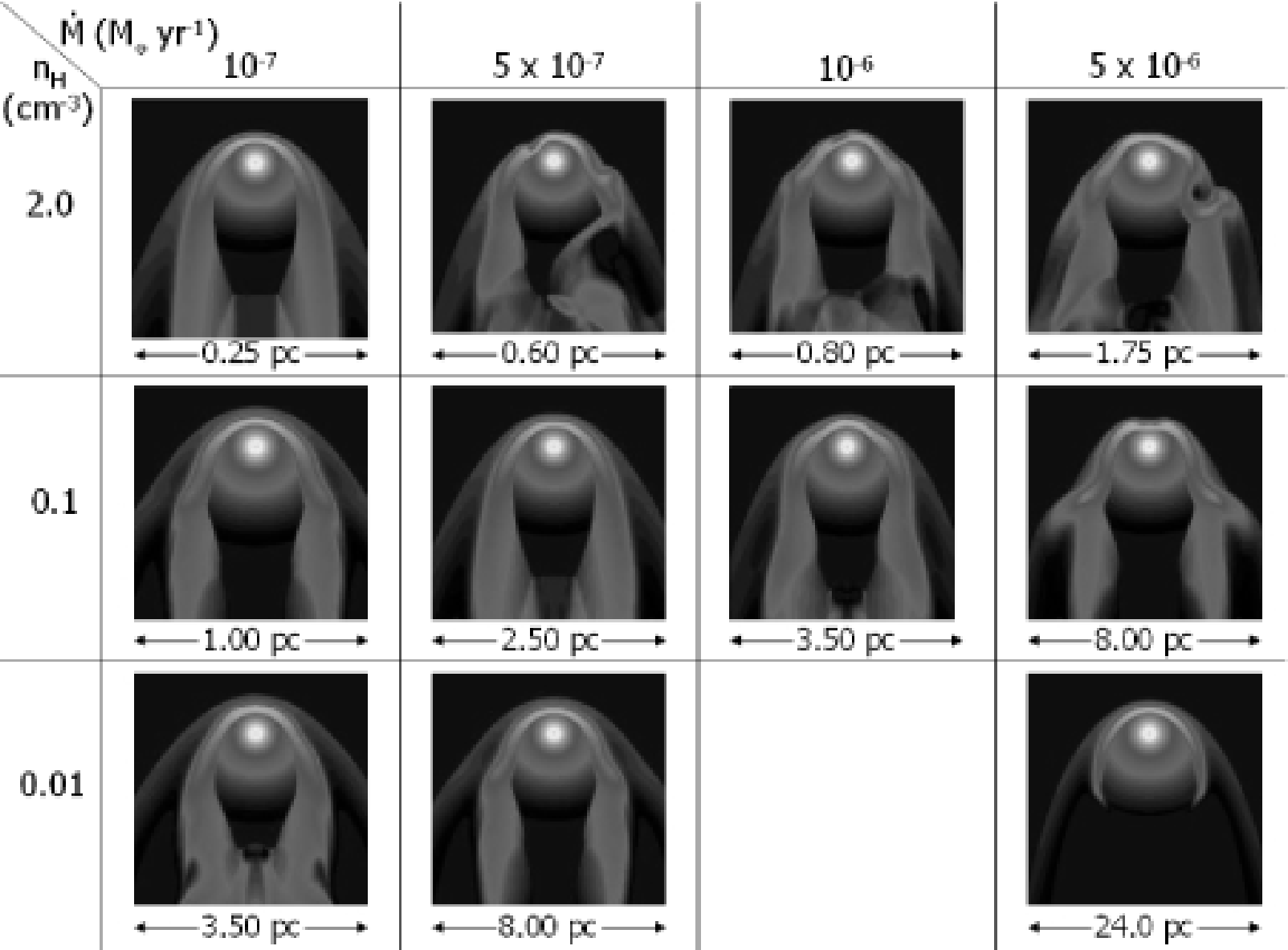}
\caption{Snapshots at the end of the AGB phase of evolution for
\vism\ = 75 \kms. mass-loss rate increases from left to
right, ISM density decreases from top to bottom. Full details of each 
simulation can be found in table \ref{apptable-75}. The 'missing'
simulation in the bottom row indicates the reduction in the
mass-loss parameter range due to computational and time constraints.}
\label{app-75}
\end{center}
\end{figure*}

\begin{table*}
\caption{Input parameters for the PN--ISM simulations in figure \ref{app-75}: 
column a) gives the mass-loss rate in the slow wind; column b) 
the density of the surronding ISM in $n_{\rm H}\,\rm cm^{-3}$; 
column c) the relative velocity of the central star; 
column d) the grid dimension along one side.}
\label{apptable-75}
\begin{center}
\begin{tabular}{ccccc}
\hline
a) & b) & c) & d)& \\
$\dot{M}_{\rm sw}$ & $\rho_{\rm ISM}$ & \vism & Grid & \\
(M$_{\odot}$\,yr$^{-1}$) & $n_{\rm H}$(cm$^{-3}$) & (\kms) & (pc)   & \\
\hline
$10^{-7}$  & 2 & 75 & 0.25 & \\
$5\times10^{-7}$  & 2 & 75 & 0.60 & \\
$10^{-6}$  & 2 & 75 & 0.80 & \\
$5\times10^{-6}$  & 2 & 75 & 1.75 & case C\\
$10^{-7}$  & 0.1 & 75 & 1.0 & \\
$5\times10^{-7}$  & 0.1 & 75 & 2.50 & \\
$10^{-6}$  & 0.1 & 75 & 3.50 & \\
$5\times10^{-6}$  & 0.1 & 75 & 8.0 & \\
$10^{-7}$  & 0.01 & 75 & 3.50 & \\
$5\times10^{-7}$  & 0.01 & 75 & 8.0 & \\
$5\times10^{-6}$  & 0.01 & 75 & 24.0 & \\
\hline
\end{tabular}
\end{center}
\end{table*}

\begin{figure*}
\begin{center}
\includegraphics[angle=0,width=16cm]{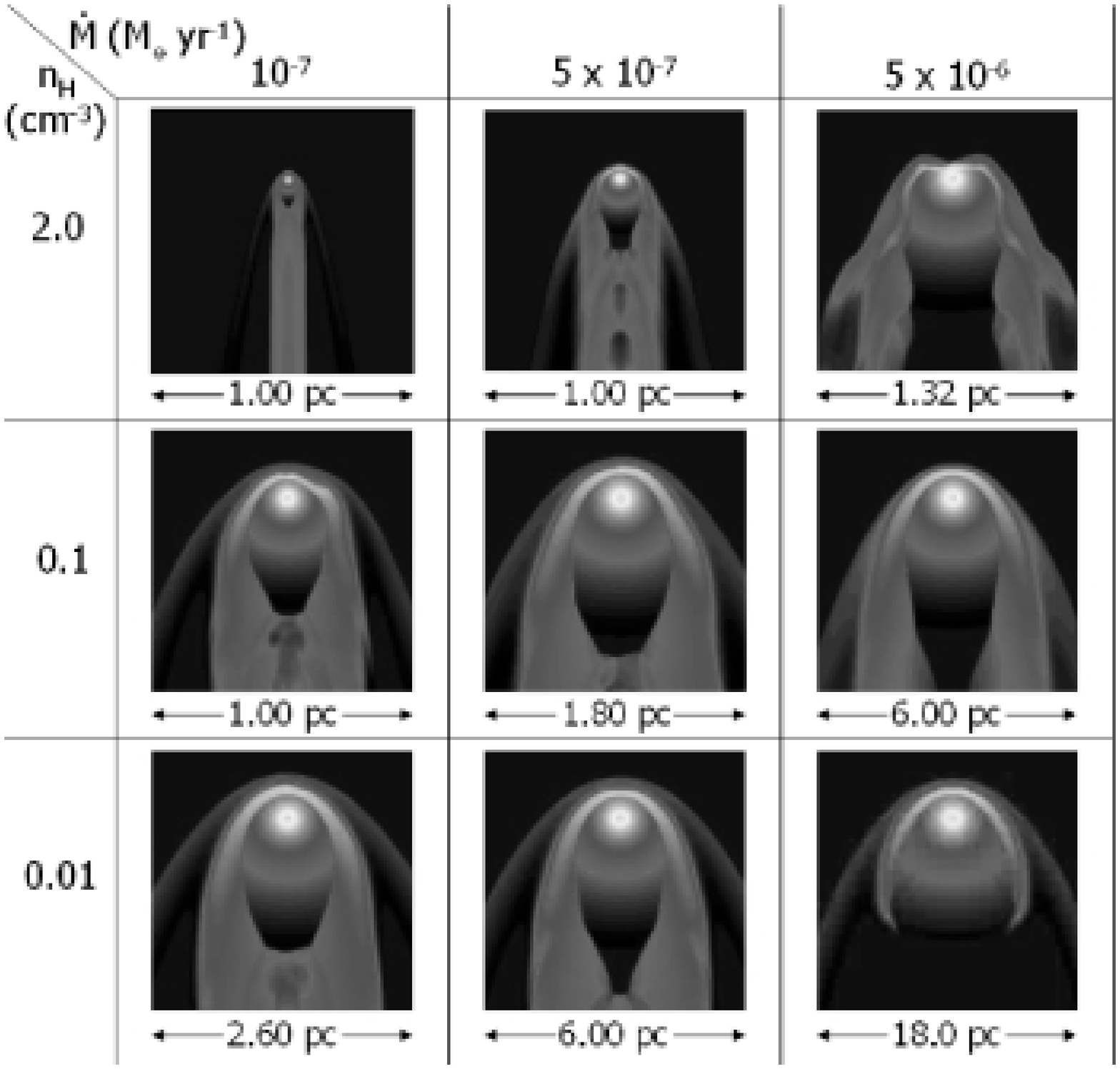}
\caption{Snapshots at the end of the AGB phase of evolution for
\vism\ = 100 \kms. mass-loss rate increases from left to
right, ISM density decreases from top to bottom. Full details of each 
simulation can be found in table \ref{apptable-100}.}
\label{app-100}
\end{center}
\end{figure*}

\begin{table*}
\caption{Input parameters for the PN--ISM simulations in figure \ref{app-100}: 
column a) gives the mass-loss rate in the slow wind; column b) 
the density of the surronding ISM in $n_{\rm H}\,\rm cm^{-3}$; 
column c) the relative velocity of the central star; 
column d) the grid dimension along one side.}
\label{apptable-100}
\begin{center}
\begin{tabular}{ccccc}
\hline
a) & b) & c) & d)& \\
$\dot{M}_{\rm sw}$ & $\rho_{\rm ISM}$ & \vism & Grid & \\
(M$_{\odot}$\,yr$^{-1}$) & $n_{\rm H}$(cm$^{-3}$) & (\kms) & (pc)   & \\
\hline
$10^{-7}$  & 2 & 100 & 2.0 & case D\\
$5\times10^{-7}$  & 2 & 100 & 2.0 & \\
$5\times10^{-6}$  & 2 & 100 & 1.32 & \\
$10^{-7}$  & 0.1 & 100 & 1.0 & \\
$5\times10^{-7}$  & 0.1 & 100 & 1.8 & \\
$5\times10^{-6}$  & 0.1 & 100 & 6.0 & \\
$10^{-7}$  & 0.01 & 100 & 2.6 & \\
$5\times10^{-7}$  & 0.01 & 100 & 6.0 & \\
$5\times10^{-6}$  & 0.01 & 100 & 18.0 & \\
\hline
\end{tabular}
\end{center}
\end{table*}

\begin{figure*}
\begin{center}
\includegraphics[angle=0,width=16cm]{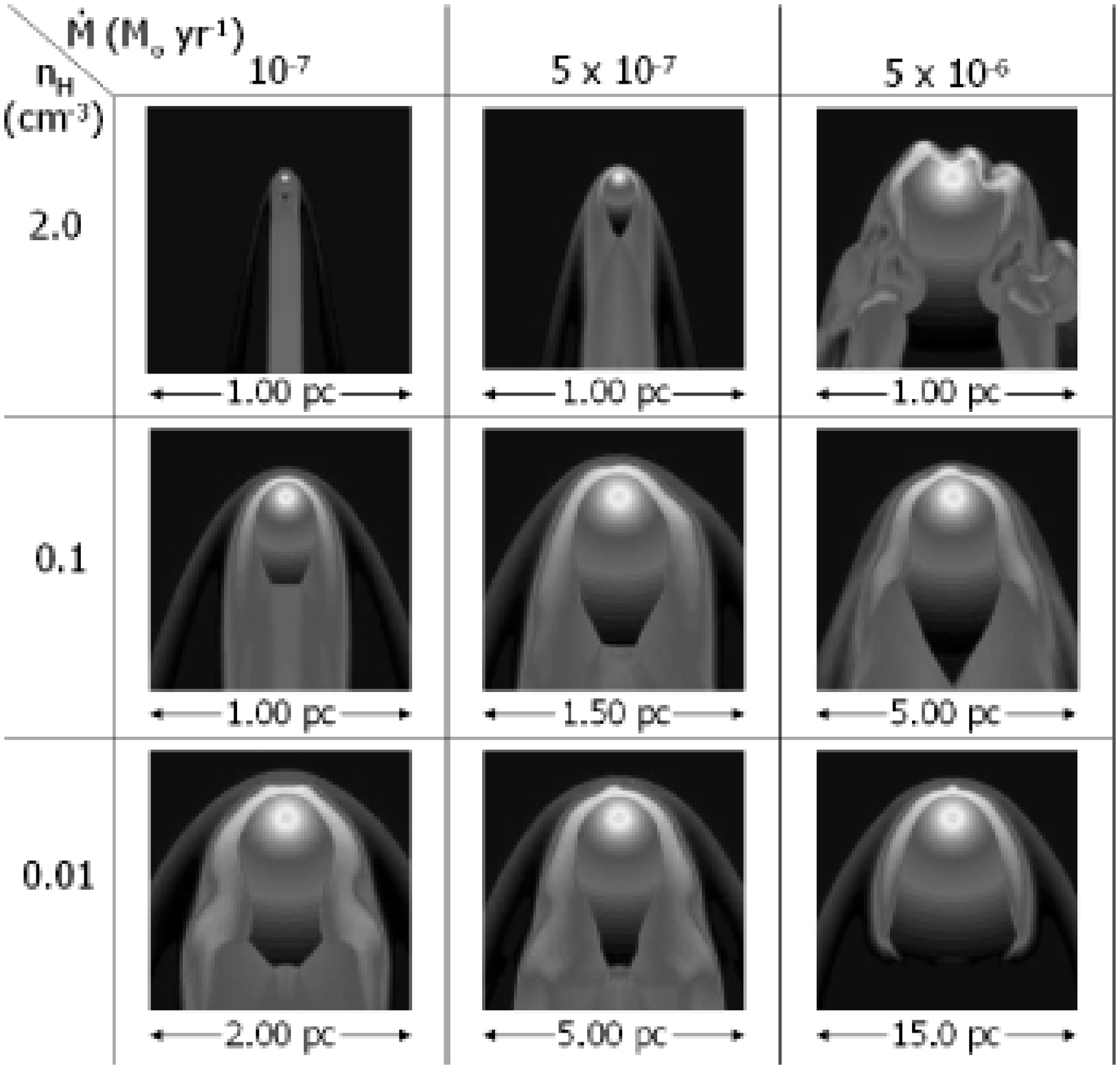}
\caption{Snapshots at the end of the AGB phase of evolution for 
\vism\ = 125 \kms. mass-loss rate increases from left to
right, ISM density decreases from top to bottom. Full details of each 
simulation can be found in table \ref{apptable-125}.}
\label{app-125}
\end{center}
\end{figure*}

\begin{table*}
\caption{Input parameters for the PN--ISM simulations in figure \ref{app-125}: 
column a) gives the mass-loss rate in the slow wind; column b) 
the density of the surronding ISM in $n_{\rm H}\,\rm cm^{-3}$; 
column c) the relative velocity of the central star; 
column d) the grid dimension along one side.}
\label{apptable-125}
\begin{center}
\begin{tabular}{ccccc}
\hline
a) & b) & c) & d)& \\
$\dot{M}_{\rm sw}$ & $\rho_{\rm ISM}$ & \vism & Grid & \\
(M$_{\odot}$\,yr$^{-1}$) & $n_{\rm H}$(cm$^{-3}$) & (\kms) & (pc)   & \\
\hline
$10^{-7}$  & 2 & 125 & 1.0 & \\
$5\times10^{-7}$  & 2 & 125 & 1.0 & \\
$5\times10^{-6}$  & 2 & 125 & 1.0 & case E\\
$10^{-7}$  & 0.1 & 125 & 1.0 & \\
$5\times10^{-7}$  & 0.1 & 125 & 1.5 & \\
$5\times10^{-6}$  & 0.1 & 125 & 5.0 & \\
$10^{-7}$  & 0.01 & 125 & 2.0 & \\
$5\times10^{-7}$  & 0.01 & 125 & 5.0 & \\
$5\times10^{-6}$  & 0.01 & 125 & 15.0 & \\
\hline
\end{tabular}
\end{center}
\end{table*}

\begin{figure*}
\begin{center}
\includegraphics[angle=0,width=16cm]{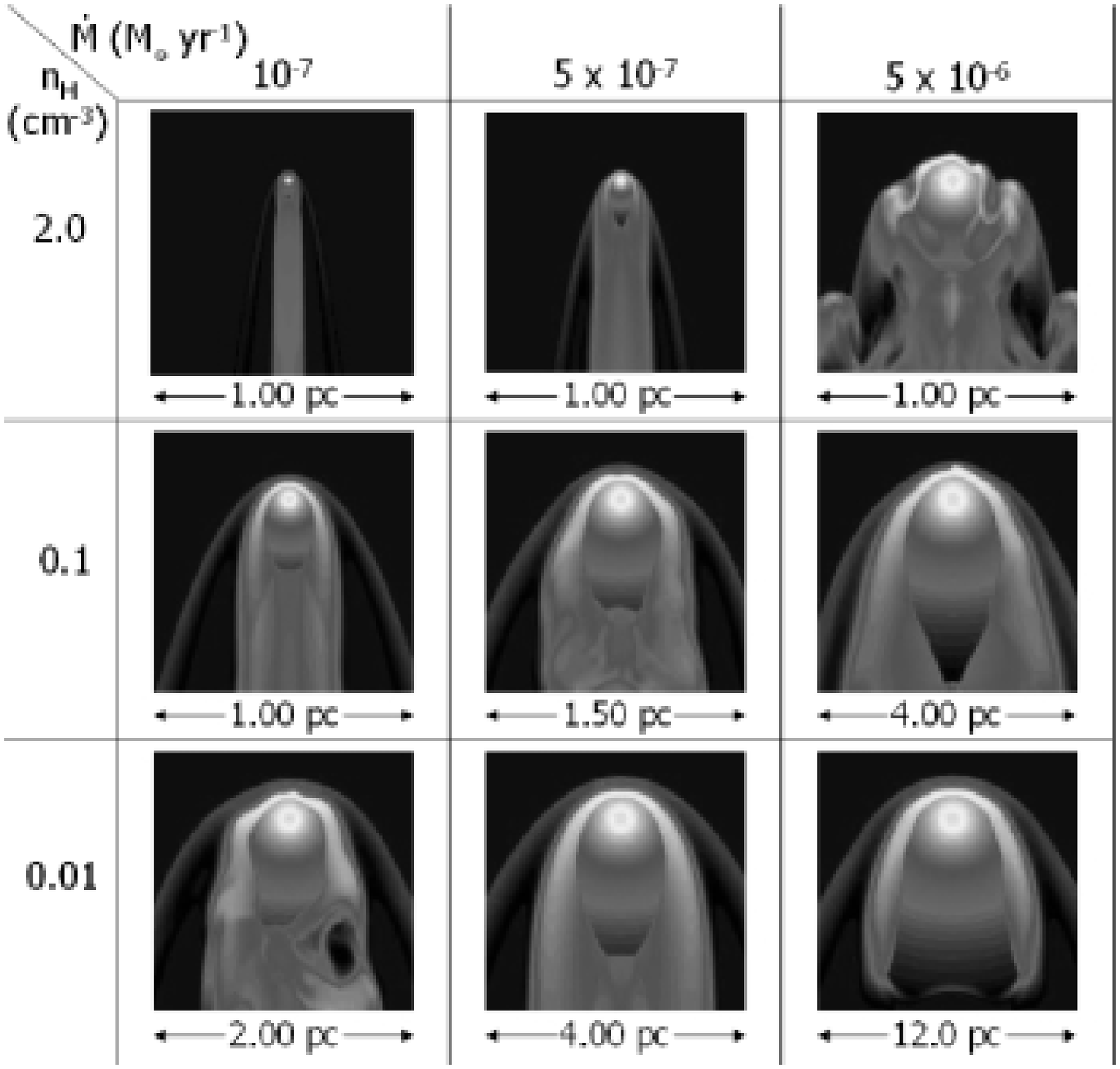}
\caption{Snapshots at the end of the AGB phase of evolution for
\vism\ = 150 \kms. mass-loss rate increases from left to
right, ISM density decreases from top to bottom. Full details of each 
simulation can be found in table \ref{apptable-150}.}
\label{app-150}
\end{center}
\end{figure*}

\begin{table*}
\caption{Input parameters for the PN--ISM simulations in figure \ref{app-150}: 
column a) gives the mass-loss rate in the slow wind; 
column b) the density of the surronding ISM in $n_{\rm H}\,\rm cm^{-3}$; 
column c) the relative velocity of the central star; 
column d) the grid dimension along one side.}
\label{apptable-150}
\begin{center}
\begin{tabular}{ccccc}
\hline
a) & b) & c) & d)& \\
$\dot{M}_{\rm sw}$ & $\rho_{\rm ISM}$ & \vism & Grid & \\
(M$_{\odot}$\,yr$^{-1}$) & $n_{\rm H}$(cm$^{-3}$) & (\kms) & (pc)   & \\
\hline
$10^{-7}$  & 2 & 150 & 1.0 & \\
$5\times10^{-7}$  & 2 & 150 & 1.0 & \\
$5\times10^{-6}$  & 2 & 150 & 1.0 & \\
$10^{-7}$  & 0.1 & 150 & 1.0 & \\
$5\times10^{-7}$  & 0.1 & 150 & 1.5 & \\
$5\times10^{-6}$  & 0.1 & 150 & 4.0 & \\
$10^{-7}$  & 0.01 & 150 & 2.0 & \\
$5\times10^{-7}$  & 0.01 & 150 & 4.0 & \\
$5\times10^{-6}$  & 0.01 & 150 & 12.0 & \\
\hline
\end{tabular}
\end{center}
\end{table*}

\begin{figure*}
\begin{center}
\includegraphics[angle=0,width=16cm]{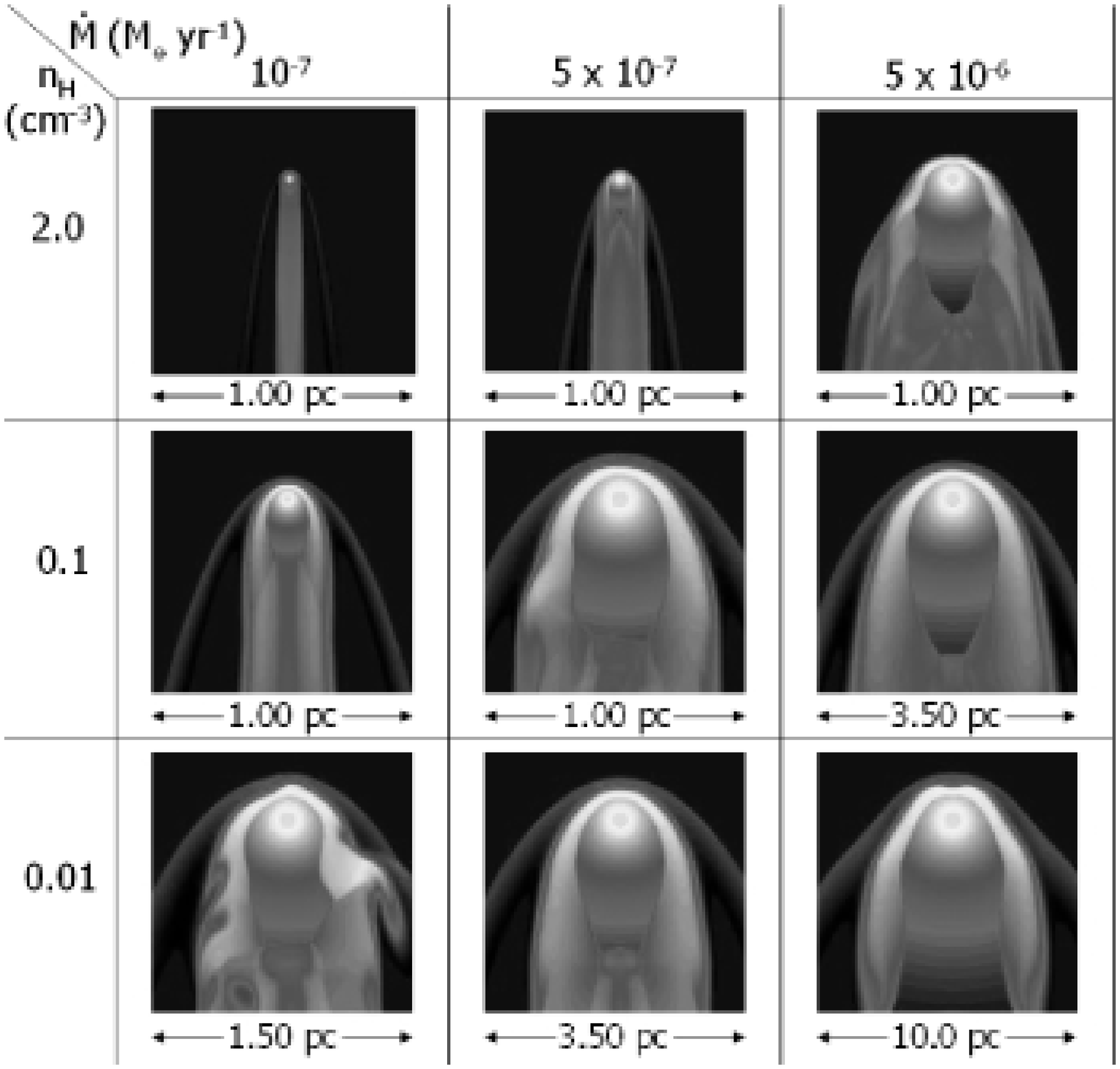}
\caption{Snapshots at the end of the AGB phase of evolution for
\vism\ = 175 \kms. mass-loss rate increases from left to
right, ISM density decreases from top to bottom. Full details of each 
simulation can be found in table \ref{apptable-175}.}
\label{app-175}
\end{center}
\end{figure*}

\begin{table*}
\caption{Input parameters for the PN--ISM simulations in figure \ref{app-175}: 
column a) gives the mass-loss rate in the slow wind; column b) 
the density of the surronding ISM in $n_{\rm H}\,\rm cm^{-3}$; 
column c) the relative velocity of the central star; 
column d) the grid dimension along one side.}
\label{apptable-175}
\begin{center}
\begin{tabular}{ccccc}
\hline
a) & b) & c) & d)& \\
$\dot{M}_{\rm sw}$ & $\rho_{\rm ISM}$ & \vism & Grid & \\
(M$_{\odot}$\,yr$^{-1}$) & $n_{\rm H}$(cm$^{-3}$) & (\kms) & (pc)   & \\
\hline
$10^{-7}$  & 2 & 175 & 1.0 & \\
$5\times10^{-7}$  & 2 & 175 & 1.0 & \\
$5\times10^{-6}$  & 2 & 175 & 1.0 & \\
$10^{-7}$  & 0.1 & 175 & 1.0 & \\
$5\times10^{-7}$  & 0.1 & 175 & 1.0 & \\
$5\times10^{-6}$  & 0.1 & 175 & 3.5 & \\
$10^{-7}$  & 0.01 & 175 & 1.5 & \\
$5\times10^{-7}$  & 0.01 & 175 & 3.5 & \\
$5\times10^{-6}$  & 0.01 & 175 & 10.0 & \\
\hline
\end{tabular}
\end{center}
\end{table*}

\begin{figure*}
\begin{center}
\includegraphics[angle=0,width=16cm]{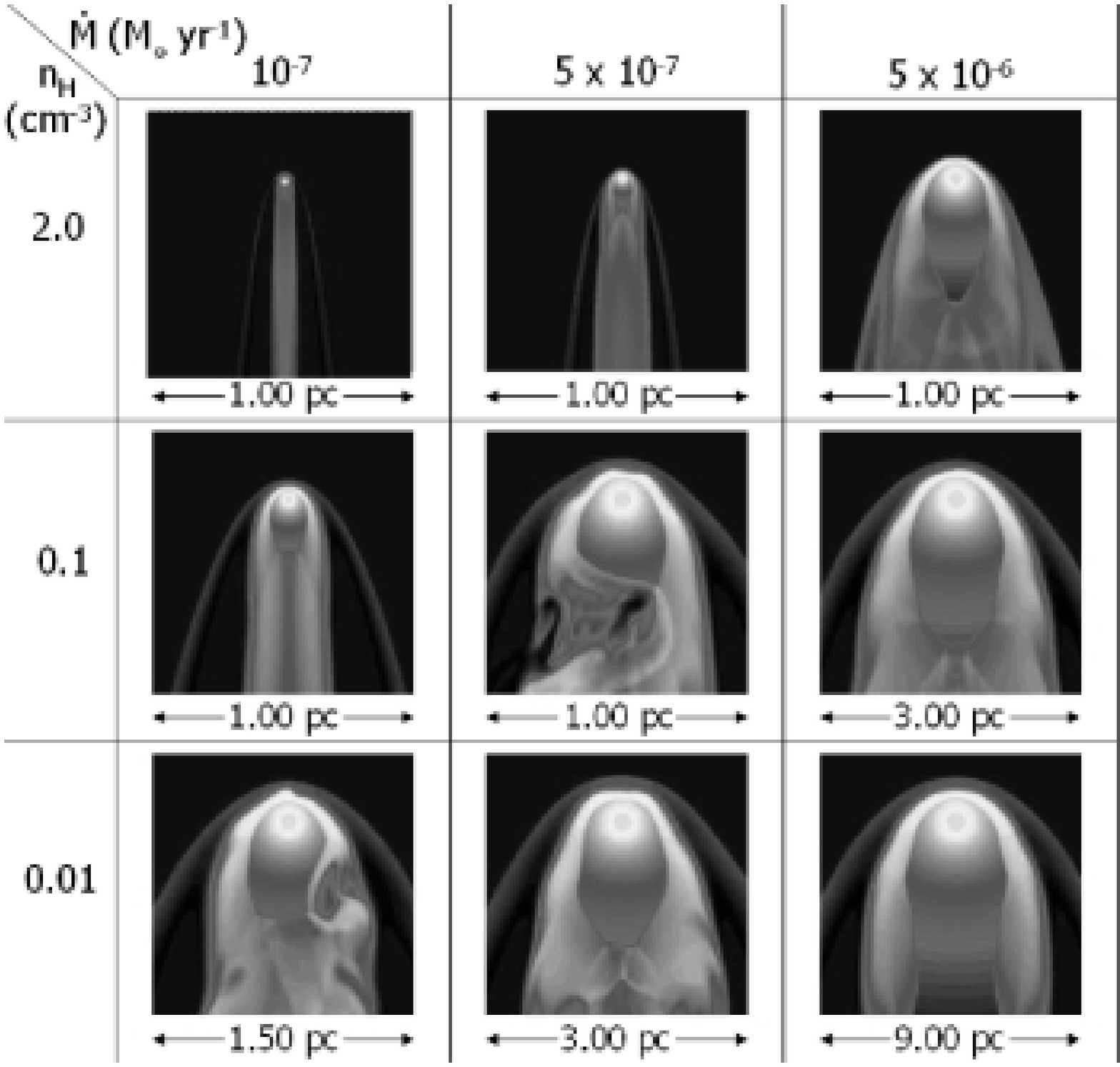}
\caption{Snapshots at the end of the AGB phase of evolution for
\vism\ = 200 \kms. mass-loss rate increases from left to
right, ISM density decreases from top to bottom. Full details of each 
simulation can be found in table \ref{apptable-200}.}
\label{app-200}
\end{center}
\end{figure*}

\begin{table*}
\caption{Input parameters for the PN--ISM simulations in figure \ref{app-200}: 
column a) gives the mass-loss rate in the slow wind; column b) 
the density of the surronding ISM in $n_{\rm H}\,\rm cm^{-3}$; 
column c) the relative velocity of the central star; 
column d) the grid dimension along one side.}
\label{apptable-200}
\begin{center}
\begin{tabular}{ccccc}
\hline
a) & b) & c) & d)& \\
$\dot{M}_{\rm sw}$ & $\rho_{\rm ISM}$ & \vism & Grid & \\
(M$_{\odot}$\,yr$^{-1}$) & $n_{\rm H}$(cm$^{-3}$) & (\kms) & (pc)   & \\
\hline
$10^{-7}$  & 2 & 200 & 1.0 & \\
$5\times10^{-7}$  & 2 & 200 & 1.0 & \\
$5\times10^{-6}$  & 2 & 200 & 1.0 & \\
$10^{-7}$  & 0.1 & 200 & 1.0 & \\
$5\times10^{-7}$  & 0.1 & 200 & 1.0 & \\
$5\times10^{-6}$  & 0.1 & 200 & 3.0 & \\
$10^{-7}$  & 0.01 & 200 & 1.5 & \\
$5\times10^{-7}$  & 0.01 & 200 & 3.0 & \\
$5\times10^{-6}$  & 0.01 & 200 & 9.0 & \\
\hline
\end{tabular}
\end{center}
\end{table*}

\end{appendix}

\label{lastpage}

\end{document}